\documentclass[pre,twocolumn,showkeys,superscriptaddress,nofootinbib,
amsmath,amssymb,aps,longbibliography]{revtex4-2}

\usepackage{etoolbox}

\makeatletter
\AtBeginDocument{%
  \patchcmd{\widetext}
    {\vrule\@height6\p@}
    {\vrule\@height22\p@}
    {}
    {}%
}
\makeatother

\usepackage{mathtools}
\usepackage[utf8]{inputenc}
\usepackage[T1]{fontenc}
\usepackage{graphicx}
\usepackage{dcolumn}
\usepackage{bm}
\usepackage{multirow}
\usepackage{array}
\usepackage{tabularx}
\usepackage{makecell}
\usepackage[table]{xcolor}
\usepackage{bbm}
\usepackage{empheq}
\usepackage{mathtools}
\usepackage{physics}
\usepackage{amsfonts}
\usepackage{microtype}

\definecolor{cblue}{rgb}{0.16,0.32,0.75}
\definecolor{cred}{rgb}{0.70,0.11,0.11}
\usepackage[
  colorlinks=true,
  linkcolor=black,
  anchorcolor=black,
  citecolor=black,
  urlcolor=black,
  bookmarks=true,
  bookmarksopen=true,
  bookmarksnumbered=true
]{hyperref}

\renewcommand{\Im}{\operatorname{Im}}

\newcommand{\E}{\mathbb{E}}
\providecommand{\Tr}{}
\renewcommand{\Tr}{\operatorname{Tr}}
\providecommand{\dd}{}
\renewcommand{\dd}{\mathrm{d}}
\newcommand{\fB}{\mathfrak{B}}
\newcommand{\GA}{\mathsf{G}_A}
\newcommand{\GNA}{\mathsf{G}_A^{(N_B)}}
\newcommand{\Gfin}{G_{\mathrm{final}}}

\newcommand{\Xh}{\widehat X}
\newcommand{\sh}{\widehat s}
\newcommand{\Yh}{\widehat Y}
\newcommand{\xih}{\widehat\xi_B}
\newcommand{\cL}{{\cal L}}
\newcommand{\Psidag}{\Psi^{\ddagger}}

\newcommand{\rsc}[1]{\rho_{\mathrm{sc},#1}}      
\newcommand{\rD}[1]{\rho_{\Delta,#1}}            
\newcommand{\Gsc}[1]{G_{\mathrm{sc},#1}}         
\newcommand{\GDl}[1]{G_{\Delta,#1}}
\newcommand{\Kk}{\mathsf{K}}                     

\def\oper{{\mathchoice{\rm 1\mskip-4mu l}{\rm 1\mskip-4mu l}
		{\rm 1\mskip-4.5mu l}{\rm 1\mskip-5mu l}}}

\begin{document}
\setcounter{secnumdepth}{3}
\title{Beyond the lemon: \\ Spectra of random Lindblad operators with a Hamiltonian component}

	
\author{Mykhailo Hontarenko}
\affiliation{Institute of Theoretical Physics, Jagiellonian University, 30--348 Cracow, Poland}
\affiliation{Doctoral School of Exact and Natural Sciences, Jagiellonian University, ul.\ Łojasiewicza 11, 30--348 Kraków, Poland}

\author{Wojciech Tarnowski}
\affiliation{Institute of Theoretical Physics, Jagiellonian University, 30--348 Cracow, Poland}

\author{Sergey Denisov}
\email{sergiyde@oslomet.no}
\affiliation{Department of Computer Science, Oslo Metropolitan University, N--0130 Oslo, Norway}

\author{Karol \.Zyczkowski}
\affiliation{Institute of Theoretical Physics, Jagiellonian University, 30--348 Cracow, Poland}
\affiliation{Center for Theoretical Physics, Polish Academy of Sciences, 02--668 Warsaw, Poland}

\date{October 23, 2026}
    
\begin{abstract}
%
Generators of continuous-time quantum Markovian evolution, known as Lindblad operators, play a key role in the theory of open quantum systems. The question of how to define and characterize random Lindblad operators has attracted considerable attention in recent years. 
Purely dissipative random Lindbladians have been extensively studied using toolboxes of free probability and non-Hermitian random matrix theories, which helped 
to uncover a characteristic lemon-like spectral support. 
In this work, we focus on random Lindbladians with a Hamiltonian component. We analyze three ensembles of increasing complexity, designed to illustrate the joint effects of dissipative and Hamiltonian dynamics. While 
two particular ensembles allow for an exact analytic treatment with standard probabilistic and scalar-valued free-probability methods, the most general case 
requires the more general framework of operator-valued free probability.
We  explain how the spectral support changes with the relative Hamiltonian strength and the rank of the dissipative component and characterize the effect of Hamiltonian-induced spectral squeezing.
\end{abstract}
\maketitle	

\section{Introduction}\label{sec:introduction}

The concept of continuous-time Markovian evolution of open quantum systems, with its natural habitats 
such as quantum optics~\cite{gardiner2004} and cavity QED~\cite{haroche2006}, has recently attracted attention as a practical tool for modelling and analyzing the evolution of noisy intermediate-scale quantum (NISQ) computational platforms~\cite{luitz2021PRR,samachLindbladTomography2022,vandenbergProbabilistic2023,brandMarkovianNoise2024,vandenbergTechniques2024,malekakhlaghEfficientLindblad2025,aguiar2025,vandenbergLargeScale2025,severinLearningLindblad2026}. In parallel, the mathematical physics community has recently shown renewed interest in different aspects of such evolution, especially its spectral properties ~\cite{baumgartner2008general,bertini2024trace,fagnola2025spectralgap,fagnola2025spectral}. In this context, the idea of a generic, non-specific random quantum Markovian evolution becomes relevant.

When the continuous-time evolution $\dot{\rho}=\mathcal{L}\rho$ of the system density operator \(\rho\) is Markovian, completely positive, trace preserving, and homogeneous in time, it constitutes a dynamical semigroup~\cite{alicki2007}, with  
all its properties  encoded in its 
generator $\cal L$. The latter can be represented in the celebrated Gorini--Kossakowski--Lindblad--Sudarshan (GKLS) form~\cite{goriniCompletelyPositiveDynamical1976,lindbladGeneratorsQuantumDynamical1976}, in which  all involved operators are rescaled such that $\hbar=1$,
\begin{align}
\label{eq:lindblad1}
\cL(\rho)
&= \cL_{\mathrm{H}}(\rho)+\cL_{\mathrm{D}}(\rho)=\\
&= -i [H,\rho] + \sum_{m,n=1}^{N^2-1} K_{m,n}\left(
F_{m}\rho F_{n}^{\dagger}
-\frac{1}{2}
\bigl\{
F_{n}^{\dagger}F_{m},\rho
\bigr\}
\right). 
\nonumber
\end{align}
Here $N$ is the dimension of the Hilbert space, $\bigl\{\cdot,\cdot \bigr\}$ is the anticommutator 
and the operators $\{F_m\}_{m=1,...,N^2-1}$ form a traceless Hilbert-Schmidt orthonormal basis, $\mathrm{Tr}\,F_m=0$, $\mathrm{Tr}(F_nF_m^{\dagger})=\delta_{mn}$~\cite{Wolf2012}. 
Henceforth, we refer to $\mathcal{L}$ as `Lindblad operator' or simply `Lindbladian'.

Once the basis is fixed, a Lindbladian is completely specified, as an operator, by two matrices, the $N\times N$ Hamiltonian matrix $H$, with  $H^\dagger=H$, and the $(N^2-1)\times(N^2-1)$ positive-semidefinite Kossakowski matrix $K$, with $K\geqslant 0$. Thus, if one asks ``What is a random Lindbladian?'', a natural answer is that it is an operator of the form 
(\ref{eq:lindblad1}), in which $H$ and $K$ are drawn at random (but still subject to the constraints of Hermiticity and positive semidefiniteness, respectively). To complete the answer, we must also specify  normalizations of $H$ and $K$, which determine the relative weights of the two components, $\cL_{\mathrm{H}}$ and $\cL_{\mathrm{D}}$~\cite{alicki2007}.

As such, random Lindbladians might be of interest from the viewpoint of random linear operator theory, especially of its part which focuses on the spectral properties of these mathematical objects~\cite{skorokhod1984,pastur1992}. However, their spectral properties are much more important from the physical point of view. Lindbladian eigenvalues, together with the corresponding eigenvectors, encode information about relaxation towards the asymptotic state, mixing, and the global structure of evolution~\cite{temme2010,macieszczak2016,saComplexSpacingRatios2020a,chruscinski2021relaxation}. In this respect, the study of random Lindbladians and their spectra~\cite{denisovUniversalSpectraRandom2019,CanOganesyan2019,akemannUniversalSignatureIntegrability2019,SaRibeiroProsen2020,langeRandomMatrixTheory2021,Tarnowski2023,yang2024decoherence} generalizes the study of random Hamiltonians and their spectra~\cite{haakeQuantumSignaturesChaos2019} and provides one of the foundations of the emerging theory of \textit{dissipative quantum chaos}~\cite{sa2026whatwetalk}.

Random Lindblad operators are natural objects to study using the mathematical apparatus of random matrix theory (RMT)~\cite{andersonGuionnetZeitouni2010}. Using vectorization of the density operator~\cite{Chruscinski2022}, these operators can be represented as $N^2\times N^2$ matrices and then can be analyzed using RMT methods, especially free probability~\cite{MingoSpeicher2017}, since the resulting matrices are non-normal.

\begin{figure*}[t]
    \centering
    \includegraphics[width=0.8\linewidth]{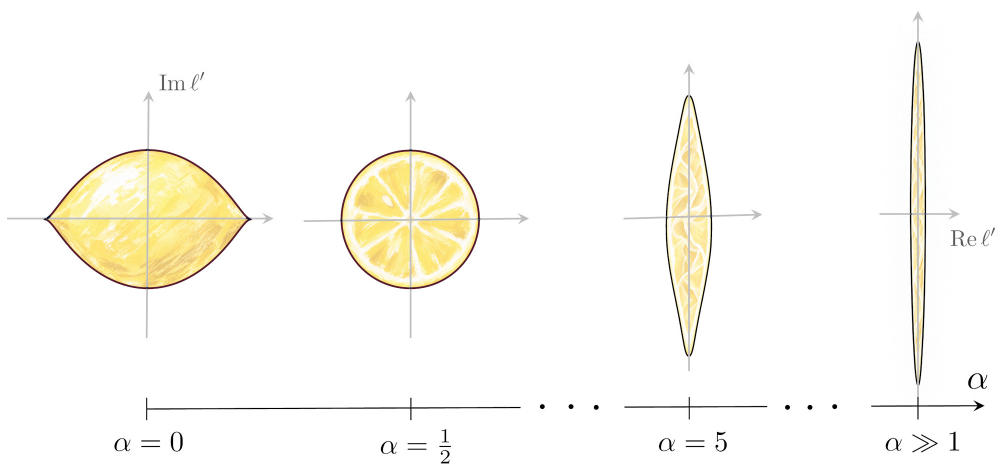}
    \caption{Rescaled spectral support $N\cdot \mathrm{spec}(\cL_{\alpha}+1)$ of a generic random Lindbladian $\cL_{\alpha}$, Eq.~(\ref{eq:lindblad4}), as a geometric set - valued function of the weight $\alpha$ of the Hamiltonian component. In the limit $\alpha=0$ it has a distinctive lemon-like shape~\cite{denisovUniversalSpectraRandom2019}.
    For $\alpha=1/2$ (we refer to as the "balanced point") the support forms a disc of radius $\sqrt{1+\sqrt{2}}$ (see Section~\ref{sec:ensembleC_silver}). An increase in the value of $\alpha$ stretches the support in the imaginary direction, while in the limit $\alpha\to\infty$,
the rescaled spectral support collapses onto the
imaginary axis.}
    \label{fig:fig0}   
\end{figure*}

Concerning the spectral densities of random Lindbladians, the only analytical result known to us are  spectral moments~\cite{langeRandomMatrixTheory2021} and the spectral support of purely dissipative random Lindblad generators, i.e., the case $H=0$~\cite{denisovUniversalSpectraRandom2019}. The spectral support has a characteristic lemon-like shape (see Fig.~\ref{fig:fig0}), which was found to be universal in the sense that it re-emerges for different constructions of random Lindblad operators~\cite{CanOganesyan2019,SaRibeiroProsen2020,luitz2020PRL,can2019JPhysA,sa2022lindbladian}. The spectral density inside the support has been evaluated using free-probability methods~\cite{Tarnowski2021}.

In this work, we go beyond the limit $H=0$ and derive an analytical equation for the spectral support of generic random Lindbladians that include a Hamiltonian component. Our presentation is self-consistent and didactic as we approach the main problem gradually, by first considering two specific ensembles of random Lindblad operators and then turning to the most generic case. For the latter, we use methods of operator-valued free probability~\cite{MingoSpeicher2017,HeltonRashidiFarSpeicher2007} and derive a closed integral equation for the  spectral boundary. 
We then solve this equation numerically and relate the stability of the numerical procedure to the intrinsic properties of the corresponding RMT model.

The paper is organized as follows. In Section~\ref{sec:setting_scene}, we provide all the necessary definitions needed to specify random Lindbladians. Section~\ref{sec:lemon} outlines the case of purely dissipative random Lindblad operators. In Section~\ref{sec:beyond}, we analyze three different ensembles of random Lindbladians with Hamiltonian components. In Section~\ref{sec:ensembleC_boundary}, we derive the equation for the boundary of the spectral support; its solvable limits and numerical evaluation are discussed in Sections~\ref{sec:ensembleC_silver} and~\ref{sec:ensembleC_numerics}. The phenomenon of spectral squeezing, observed for two ensembles, is discussed in Section~\ref{sec:squeezing}
The paper concludes with Section~\ref{sec:conclusion}, which summarizes the results and outlines several open questions. Technical details and derivations are presented in the Appendices.

\section{Construction of random Lindblad operators}
\label{sec:setting_scene}
In this section we describe the construction of random Lindbladians and the methods to sample them.

We start with the
Kossakowski matrix $K$. It is a positive semi-definite and therefore 
it admits a representation $K=JJ^{\dagger}$, where $J$ is a complex-valued matrix with no other constraints. The rank of $K$, denoted by $R$, bounded by $R \leqslant N^2-1$ (and thus the number of jump operators, see below) can be controlled via rectangularity of matrix $J\in\mathbb{C}^{(N^2-1)\times R}$. Jump operators
are defined as $L_{k}=\sum_{m=1}^{N^2-1}J_{mk}F_m$,
$k=1,\ldots,R$~\cite{breuerTheoryOpenQuantum2010,Wolf2012}.  
The dissipative part now reads
\begin{equation}
\begin{aligned}
    \cL_{\mathrm{D}}(\rho)
    = \sum_{k=1}^{R} \Bigl(
    L_{k}\rho L_{k}^{\dagger}
    - \frac{1}{2}
    \{ L_{k}^{\dagger}L_{k},\rho \}
    \Bigr).
\end{aligned}
    \label{eq:lindblad2}
\end{equation}

The representation $K=JJ^{\dagger}$ suggests a way to sample $\cL_{\mathrm{D}}$ at random. Namely, we sample a rectangular complex Ginibre matrix $G\in\mathbb{C}^{(N^2-1)\times R}$ with independent and identically distributed normal entries, $g_{kl}\sim\mathcal{N}_{\mathbb C}(0,1)$.
The resulting Kossakowski matrix is then a Wishart matrix~\cite{denisovUniversalSpectraRandom2019}. Finally, we normalize $K=N \cdot GG^{\dagger}/\Tr GG^{\dagger}$,
so that $\Tr K = N$ and $J=\sqrt{\frac{N}{\Tr(GG^{\dagger})}}\cdot G$.

The dissipative part can also be represented as~\cite{Wolf2012} 
\begin{equation}
\cL_{\mathrm{D}}(\rho) =  \Psi(\rho) -\frac{1}{2}\left( \Psidag(\oper)\rho + \rho  \Psidag(\oper)\right), 
  \label{eq:lindblad3}
\end{equation}
where $\Psi(\rho) = \sum_{k=1}^{R} L_{k}\rho L_{k}^{\dagger} $ is a completely positive (CP),  but not necessarily trace preserving (TP), map of rank $R$, and $\Psidag$ is the map dual to $\Psi$. Note that jump operators $\{L_{k}\}$ become Kraus operators of the auxiliary map $\Psi$.

The representation \eqref{eq:lindblad3} offers an alternative way to construct a random ensemble of $\cL_{\mathrm{D}}$, via an ensemble of random CP maps. If the CP maps are also TP, several sampling procedures are described in Refs.~\cite{bruzdaRandomQuantumOperations2009,Kukulski2021}. Namely, one possibility is to sample a random Choi matrix $C_{\Phi}\geqslant 0$, for example from a Wishart ensemble, and then impose the TP condition, $\Tr_{2}C_{\Phi}=\oper$. Alternatively, one can sample $R$ independent random Kraus operators $\{L_{k}\}_{k=1}^{R}$ 
and impose trace preservation by the normalization $L_{k}\rightarrow L_{k}/\sqrt{\sum_{k=1}^{R}L_{k}^{\dagger}L_{k}}
$. 
Finally, one may use a Stinespring dilation~\cite{Wolf2012}, by sampling a Haar-random unitary $U\in\mathrm{U}(NR)$ (or, more economically,  a random isometry $V:\mathbb{C}^N \rightarrow \mathbb{C}^N\otimes\mathbb{C}^R,
~V^\dagger V=\oper$) acting on the system and an $R$-dimensional ancilla, fixing an ancilla state $|0\rangle \langle0|$, and defining $\Phi(\rho)=\Tr_{\mathrm{a}}\!\left[U(\rho\otimes|0\rangle\langle0|)U^{\dagger}\right]$. Note that in this case the Kraus operators are $L_{k}=\langle k|U|0\rangle$. 

However, in our case the map $\Psi$ need not be TP (imposing this condition leads to a more restricted Ensemble B, see Section~\ref{ensembleB}). Without the TP constraint, sampling via a Choi matrix or a random set of Kraus operators becomes simpler and straightforward. In the Stinespring construction, the isometry should be replaced by an arbitrary linear map
$A:\mathbb{C}^N \rightarrow \mathbb{C}^N \otimes \mathbb{C}^R$.

The normalizations corresponding to $\Tr K=N$ are implemented as follows. The Choi matrix has to be normalized as
$C_\Phi \rightarrow 
N \cdot \,C_\Phi/[
{\Tr C_\Phi-\frac{1}{N}
\langle\oper_{N^2}|C_\Phi|\oper_{N^2}\rangle}]$,
while the Kraus operators are normalized as
$L_k\rightarrow \sqrt{\frac{N}{S_K}}\,L_k$,
where
$S_K=\sum_{k}\Tr\!\left( L_k^\dagger L_k\right)$.
Finally, for Stinespring-based sampling, we should first derive the corresponding set of Kraus operators,
$L_k=(\oper\otimes\langle k|)A$,
and then normalize them as before.

The Hamiltonian component, $\cL_{\mathrm{H}}(\cdot)=-i[H,\cdot]$, is determined by the Hermitian matrix $H$, $H=H^\dagger$. A natural choice is to sample $H$ from the Gaussian unitary ensemble (GUE)~\cite{andersonGuionnetZeitouni2010},  with probability density
$P(H)\propto \exp[-\frac{N}{2}\Tr H^2]$.
With this convention, $\mathbb{E}\Tr H^2=N$ and the eigenvalue density converges, in the limit $N\to\infty$, to the Wigner semicircle $\rsc{1}(x)$ (henceforth we refer to the Wigner circle supported on the interval $[-a,a]$ as $\rsc{a}(x) \coloneqq \frac{2}{\pi a^{2}}\sqrt{a^{2}-x^{2}}$). We impose the normalization $\Tr H^2=1/N$~\cite{denisovUniversalSpectraRandom2019}, so each realization is rescaled as $H\longrightarrow H/\sqrt{N\Tr H^2}$.

The particular choice of the normalizations for $K$ and $H$ is aimed at balancing contributions of the corresponding components to the spectrum of $\cL$.
In the case of full rank, the Kossakowski matrix $K$ has dimension $(N^2-1) \times (N^2-1)$. With $\Tr K=N$, its mean eigenvalue is $\Tr K/(N^2-1)=N/(N^2-1)\simeq 1/N$. After all $N^2-1$ dissipative channels are combined, the Lindbladian spectrum is centered at a value of order unity, $\langle \ell_{\mathrm D}\rangle \simeq -1$. Indeed, $\Tr_{\mathrm{HS}}\cL_{\mathrm D}=-N\Tr K=-N^2$, so the mean nonzero Lindbladian eigenvalue approaches $-1$.  In the large-$N$ limit, $\cL_{\mathrm D}\simeq -\oper+\frac{1}{N}\cL'_{\mathrm D}$, and so the width of the spectral cloud scales as $1/N$.
The spectrum of the Hamiltonian part $\cL_{\mathrm H}=-i[H,\cdot]$ is centred at zero. For $\Tr H^2=1/N$,  the eigenvalues of $H$ are of order $1/N$, and hence the differences $E_j-E_k$ are also of order $1/N$. Therefore, the width of the Hamiltonian component spectrum is $\operatorname{width}\operatorname{spec}(\cL_{\mathrm H})=O(1/N)$. As a result, we have
\begin{equation}
\operatorname{width}[\operatorname{spec}
\left(\cL_{\mathrm D}+\oper\right)]
\sim 
\operatorname{width}[\operatorname{spec}
\left(\cL_{\mathrm H}\right)]
\sim \frac{1}{N}.
\end{equation}

Now we can define the relative weight $\alpha$ of the Hamiltonian component with respect to the dissipative one. That is, if the random Hamiltonian $H$ is sampled and normalized as above, we use an effective Hamiltonian $H_{\mathrm{eff}}=\alpha H$. In other words, if the Hamiltonian and Kossakowski components are normalized as discussed above, we are interested in the parametrized family of  Lindblad generators,
\begin{equation}
\label{eq:lindblad4}
\cL_{\alpha}(\rho)= 
-i\alpha [H,\rho] + 
\cL_{\mathrm{D}} (\rho) \ .
\end{equation}

Our aim is to derive the boundary of the spectral support (spectral contour) for the Lindblad generators (\ref{eq:lindblad4})
as a function of the real parameter $\alpha \geq 0$; see Fig.~\ref{fig:fig0}.

To obtain the spectrum $\{\ell_j\}_{j=0,...,N^2-1}$ of a Lindblad operator numerically, the latter has to be recast in a matrix form using, e.g., the column-stacking convention~\cite{Chruscinski2022},
$|\rho\rangle\rangle=\operatorname{vec}(\rho)$,
for which
$
\operatorname{vec}(A\rho B)
=
(B^T\otimes A)\operatorname{vec}(\rho).
$
Thus, a linear map $\mathcal{E}$ is represented by an $N^2\times N^2$ matrix $\widehat{\mathcal{E}}$ such that
$
|\mathcal{E}(\rho)\rangle\rangle
=
\widehat{\mathcal{E}}|\rho\rangle\rangle
$.

For the Hamiltonian part of the Lindbladian we have
~$\widehat{\mathcal L}_H =
-i\left(\oper\otimes H -
\bar{H}\otimes \oper \right)$.
The dissipative part in the representation~\eqref{eq:lindblad1} can be written as
\begin{equation*}
\begin{aligned}
\widehat{\mathcal L}_{\mathrm{D}}
&=
\sum_{a,b=1}^{N^2-1}
K_{ab}
\Biggl[
\overline{F_b}\otimes F_a
-
\frac12
\Bigl(
\oper \otimes F_b^\dagger F_a
+
(F_b^\dagger F_a)^T\otimes \oper
\Bigr)
\Biggr].
\end{aligned}
\end{equation*}
For the jump-operator representation~\eqref{eq:lindblad2}, column stacking gives the following $N^2\times N^2$ matrix:
\begin{equation*}
\begin{aligned}
\widehat{\mathcal L}_{\mathrm{D}}
&=
\sum_k
\Biggl[
\overline{L_k}\otimes L_k
-
\frac12
\Bigl(
\oper\otimes L_k^\dagger L_k
+
(L_k^\dagger L_k)^T\otimes \oper
\Bigr)
\Biggr].
\end{aligned}
\end{equation*}
For an auxiliary CP map
$\Psi(\rho)=\sum_k L_k\rho L_k^\dagger, $ the vectorized form is
$
\widehat{\Psi}
=
\sum_k
\overline{L_k}\otimes L_k
$. This gives for the representation~\eqref{eq:lindblad3}
\begin{equation}
\label{eq:lindblad4_vector}
\widehat{\mathcal L}_{\mathrm{D}}
=
\widehat{\Psi}
-
\frac12
\left(
\oper \otimes \Psidag(\oper)
+
\Psi^{\ddagger}(\oper)^T\otimes \oper
\right).
\end{equation}
Here, $\Psi^\ddagger(\oper)=\sum_k L_k^\dagger L_k$ is a Hermitian positive-semidefinite matrix, $A=A^\dagger\geq 0$.

Note that, in any orthonormal Hermitian operator basis, $F_a^\dagger=F_a$, the Lindbladian has a real matrix representation due to Hermiticity preservation. Moreover, since $\Psi^\ddagger(\oper)$ is real symmetric and positive semidefinite, a suitable unitary rotation can bring it into a real symmetric (not necessarily diagonal) form, while preserving the reality of the Lindbladian matrix representation.

Finally, since Lindbladians preserve Hermiticity and annihilate the trace. Their spectra contain a zero eigenvalue corresponding to a stationary state and are symmetric under complex conjugation, that is if $\ell \in \mathrm{spec}(\mathcal L)$ then $\ell^* \in \mathrm{spec}(\mathcal L)$.

\section{The lemon}\label{sec:lemon}

In this section we outline the evaluation of the spectral support of purely dissipative random Lindblad generators, $\cL_D = \cL_{0}$, Eq.~(\ref{eq:lindblad4}), by RMT methods. For a more detailed description, we refer the reader to Refs.~\cite{denisovUniversalSpectraRandom2019, Tarnowski2021}. 

We start with the matrix form given by Eq.~(\ref{eq:lindblad4_vector}) and introduce matrix $X  =\Psi^\ddagger(\oper)-\oper$. This is a Hermitian matrix (since $\Psi^\ddagger$ preserves the Hermiticity) and, if $\Psi$ is trace-preserving, it equals zero (since $\Psi^\ddagger(\oper) = \oper$). Moreover, because of the normalization $\Tr K=N$, together with unitary invariance of the sampling, $\mathbb{E}\,\Psi^\ddagger(\oper)=\oper$,  $\mathbb{E}\,X=0$. Thus, Eq.~(\ref{eq:lindblad4_vector}) can be rewritten as $\widehat{\cL}_D =
\widehat{\Psi} - \oper\otimes\oper
- \frac{1}{2} \left(\oper\otimes X+
X^T\otimes\oper
\right)$. Now, assuming that the basis is Hermitian and $\Psi^\ddagger(\oper)$ is real symmetric, 
we can build the RMT model of the matrix $N(\cL_D + \oper_{N^2})$,
\begin{equation}
\hat{\cL}_{\mathrm{D}}^{RMT} \simeq  Y  -   \left(
        \oper\otimes C
        +
        \overline C\otimes\oper
    \right),
    \label{eq:lemon_RMT_dissipative}
\end{equation}
where $Y$ is an $N^2 \times N^2$  real square Ginibre matrix, i. e., a member of Ginibre Orthogonal Ensemble (GinOE), and $C$ is a member of  Gaussian Orthogonal Ensemble (GOE). In addition, the normalizations
$\operatorname{Tr}(Y Y^\dagger)=N^2$ and $
\operatorname{Tr}C^2=N/4$ are enforced. We can also sample a random matrix $C$ from Gaussian unitary ensemble  (GUE), that would not change the resulting support in large-$N$ limit. 

The term $\oper\otimes C+\overline{C}\otimes\oper$ is Hermitian and can therefore be diagonalized by a unitary $\hat V$. We relax, for the moment, the condition that $Y$ has to be real, since this does not affect the geometry of the spectral support of $\hat{\cL}_D^{\mathrm{RMT}}$ in the large-$N$ limit, and assume instead that $Y$ is a properly normalized member of the complex Ginibre ensemble (GinUE) and consider $\hat V\hat{\cL}_D^{\mathrm{RMT}}\hat V^\dagger$. Since the GinUE is invariant under unitary rotations, $\hat VY\hat V^\dagger$ is again a member of the GinUE. If $\{\mu_i\}_{i=1}^N$ are the eigenvalues of $C$, we therefore obtain
$\hat V\hat{\cL}_{\mathrm{D}}^{\mathrm{RMT}}\hat V^\dagger
=
Y-\operatorname{diag}\{\mu_j\}_{j=1}^{N^2}$,
where  $\mu_j$ are distributed according to the classical self-convolution of the Wigner semicircle distribution (see Appendix \ref{app:convolution}).

\begin{figure}[t]
    \centering
    \includegraphics[width=0.99\linewidth]{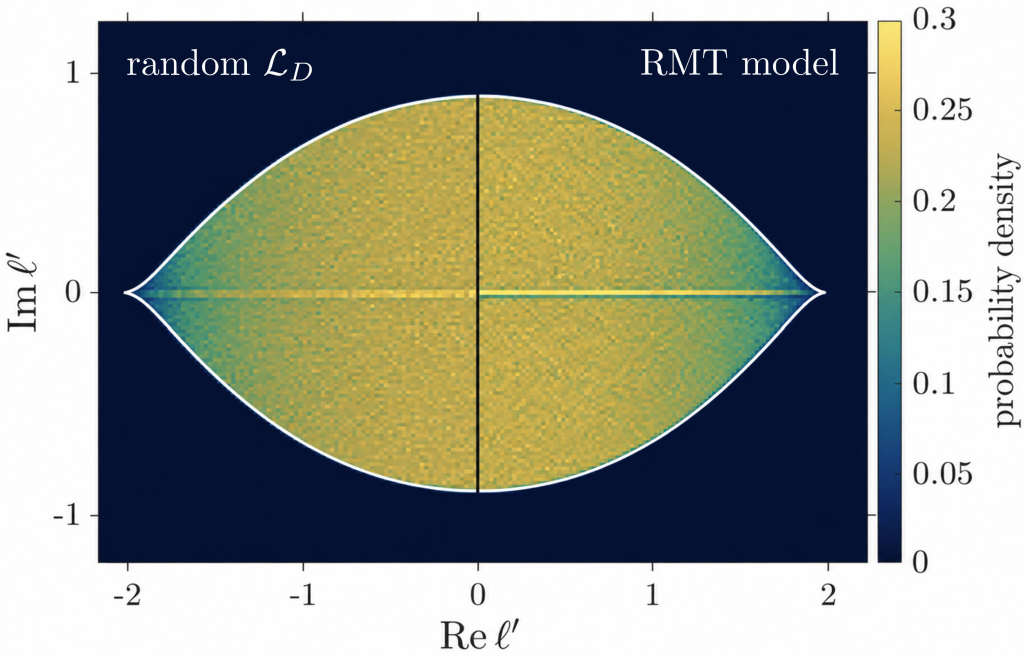}
    \caption{Spectral density of the rescaled purely dissipative random Lindbladian $N(\cL_D+\oper)$ (left) and the corresponding RMT model, Eq.~(\ref{eq:model_1}) (right), for $N=100$. The white line shows the analytical lemon-shaped boundary given by Eq.~(\ref{eq:lemon_contour}). Each distribution was sampled with $10^3$ realizations.}
\label{fig:fig_lemon}     
\end{figure}

We thus arrive at the conclusion that the spectral support can be approximated as the free sum~\cite{MingoSpeicher2017} of a circular element $Y$ and a Hermitian diagonal variable $D$ with the specific eigenvalue density (since this density is symmetric about zero, the sign of $D$ is irrelevant), 
\begin{equation}
\widehat{\mathcal L}^{\mathrm{RMT}}_D \sim Y + D.
\label{eq:model_1}
\end{equation}
The boundary of the spectral support can be evaluated analytically~\cite{denisovUniversalSpectraRandom2019}. It is a  lemon-shaped contour (see Fig.~\ref{fig:fig0}) given by the solution of the equation
\begin{equation}
\operatorname{Im}\left[z+\GDl{1}(z)\right]=0,
\label{eq:lemon_contour}
\end{equation}
where
\begin{equation*}
\begin{aligned}
\GDl{1}(z)
    &=
    2z
    -
    \frac{2z}{3\pi}
    \Biggl[
        (4+z^2)
    \mathrm{E}\!\left(\frac{4}{z^2}\right)+
    (4-z^2)
    \mathrm{K}\!\left(\frac{4}{z^2}\right)
\Biggr].
\end{aligned}
\end{equation*}
Here, $\mathrm{K}(x)$ and $\mathrm{E}(x)$ denote the complete elliptic integrals of
the first and second kind~\cite{GradshteynRyzhik}. The density inside the contour can be evaluated using the quaternionic Green's function formalism~\cite{Tarnowski2021}. The RMT model reproduces the spectral density of  random $\mathcal L_{\mathrm D}$ ensembles very accurately starting already at $N \sim 100$; see Fig.~\ref{fig:fig_lemon}.

So far we addressed the full-rank case corresponding to $R=N^2-1$. More generally, one can consider a random Kossakowski matrix of rank $R<N^2-1$. This can be realized by sampling a rectangular complex Ginibre matrix $G\in\mathbb{C}^{(N^2-1)\times R}$ and then constructing the normalized Wishart matrix
$K=N\,\frac{GG^\dagger}{\Tr(GG^\dagger)}$. If the normalization  is enforced, the center of the spectral bulk  remains at $-1$, whereas its size scales as $R^{-1/2}$. Introducing the relative rank $r=R/(N^2-1)$, one therefore has
\begin{equation*}
\operatorname{width}[\operatorname{spec}(\cL_{\mathrm{D}}+\oper)]
\sim \frac{1}{\sqrt{R}}
\simeq \frac{1}{N\sqrt{r}}.
\end{equation*}
Consequently, in the rescaled variable $\ell'=N(\ell+1)$, the lemon-shaped support is dilated by a factor $1/\sqrt{r}$ with respect to the full-rank case. Equivalently, the scaling $\ell'=\sqrt{R}\,(\ell+1)$ removes the rank dependence.

The lemon-shaped support is not specific to the  Wishart construction. Closely related spectral contours have been observed for several other ensembles of  dissipative Lindblad generators, including alternative constructions of the dissipative part~\cite{CanOganesyan2019,can2019JPhysA},  random Lindbladians with local jump operators~\cite{luitz2020PRL}, and dissipative many-body models~\cite{sa2022lindbladian}. These results highlight that the lemon contour represents a robust spectral feature rather than a property of a particular sampling recipe.

\section{Stepping beyond the lemon: Three ensembles of random Lindbladians with Hamiltonian components} \label{sec:beyond}

Now we turn to the case where the Hamiltonian component is present in a random Lindbladian with $\alpha \neq 0$ in Eq.~(\ref{eq:lindblad4}). The Hamiltonian $H$ is normalized such that $\operatorname{Tr}H^2=1/N$. 

After the rescaling,
$    \widehat{\mathcal L}_\alpha'
    =
    N
    \left(
        \widehat{\mathcal L}_\alpha
        +
        \oper_{N^2}\right)
$,
the corresponding random-matrix model can be written as
\begin{equation}
\begin{aligned}
        \widehat{\mathcal L}_\alpha^{RMT}
    &\simeq
    Y
    -
    \left(     
         \oper\otimes W        +  \overline W\otimes\oper \right) ,\\
    W&=C+i\alpha H_0.
\end{aligned}
\label{eq:RMT_general}
\end{equation}

Note that, in the RMT model, rescaling the Lindbladian by a factor of $N$ gives $H_0=NH$ and, correspondingly, $\operatorname{Tr}H_0^2=N$. $H_0$ can be modelled as a traceless GUE matrix (since the identity component does not contribute to the commutator), rescaled to satisfy the normalization.

In the following subsections, we consider three ensembles of Lindblad generators of increasing complexity to illustrate how the interplay between the Hamiltonian and dissipative components shapes the spectral density. The third ensemble, Ensemble C, corresponds to the general case, Eq.~(\ref{eq:lindblad4}).

\subsection{Ensemble A: Commuting Hamiltonian and jump operators}
\label{sec:ensembleA}
We first consider a simple model of a Lindbladian with Hamiltonian and  all jump operators diagonal in a common basis,
\begin{equation}
\begin{aligned}
 H=\operatorname{diag}(\varepsilon_1,\ldots,\varepsilon_N),~~
 L_m=\operatorname{diag} (v^m_{1},\ldots,v^m_{N}),\\~m=1,2,...,R~~~~~~~~~~~~~~~~~~~~~~~.
 \label{eq:Asetup}
\end{aligned}
\end{equation}
The dissipative part is characterized by the set of rates $\{\gamma_m\}_{m=1,2,...,R}$ and the total dephasing rate $ \Gamma_D:=\sum_{m=1}^R\gamma_m$. 

It is evident  that the steady state of the model Lindbladian belongs to the manifold of density matrices diagonal in the common eigenbasis. However, the dephasing processes with and without the Hamiltonian part are different. The toy model \eqref{eq:Asetup} highlights several important effects and provides a reference point for interpreting the results obtained for the other two ensembles.

To begin, we  assume that the diagonal jump eigenvalues \(v_{mi}\) are real and sampled from the semicircle distribution $\rsc{1}(x)$. In the limit $N,R \gg 1$, as discussed below, the particular choice of the distribution is not essential, provided that it has a finite variance. The Hamiltonian eigenvalues are sampled from the standard GUE semicircle distribution $\rho_{\mathrm{sc},2}$.

Since the matrix units $E_{ij}=|i\rangle\langle j|$
diagonalize both components of the Lindbladian, the eigenvalues of the latter are \begin{equation}
\begin{aligned}
 \ell_{ij}= h_{ij} + d_{ij} =-i\alpha(\varepsilon_i-\varepsilon_j)+~~~~~~~~~~~~~~\\
\sum_{m=1}^R\gamma_m\left[v_{mi}\overline{v_{mj}}-\frac{|v_{mi}|^2+|v_{mj}|^2}{2}\right].
 \label{eq:Aexact}
 \end{aligned}
\end{equation}
The $N$ diagonal modes $E_{ii}$ have eigenvalues $\ell_{ii}=0$ and hence the zero eigenvalue is $N$-fold degenerate.

\begin{figure*}[t]
\centering
\includegraphics[width=0.95\textwidth]{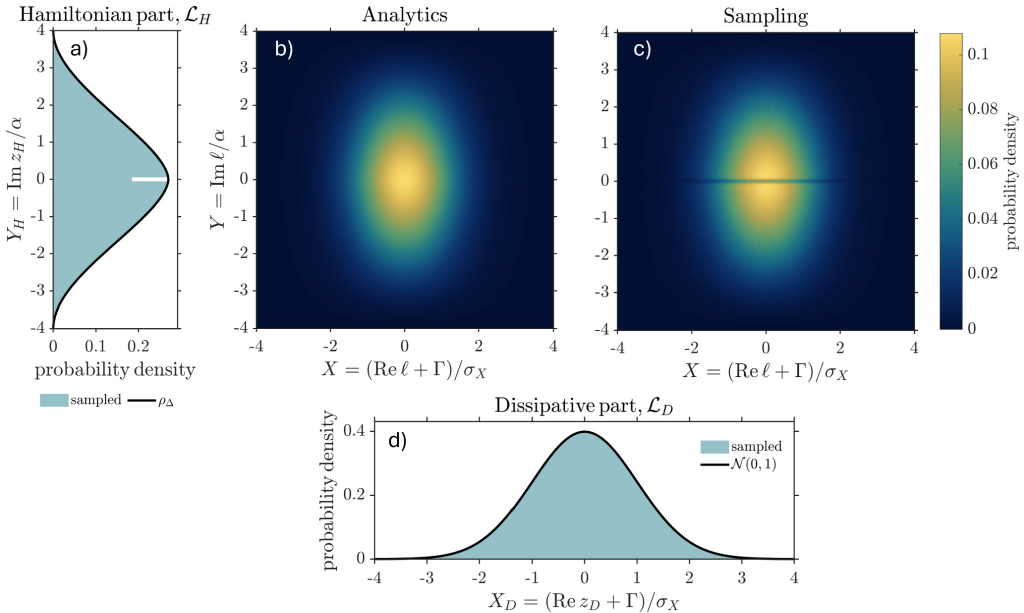}
\caption{The rescaled spectral density $\ell' = \sqrt{\frac{2R}{3}}\,\frac{\operatorname{Re}\ell+\Gamma_D}{\Gamma_D} + i\,\frac{\operatorname{Im}\ell}{\alpha}$ of Lindbladians $\cL_{\alpha}$, Eq.~(\ref{eq:lindblad4}), with commuting Hamiltonian and jump operators. The jump eigenvalues are sampled from the Wigner semicircular distribution, with equal dissipative rates $\gamma_m=\Gamma_D/R$, where $\Gamma_D=\sum_{m=1}^{R}\gamma_m$ is the total dissipative rate. The Hamiltonian is drawn from the GUE with the spectral support $[-2,2]$.
The spectral density of the dissipative part approaches a standard Gaussian distribution as $N,R\to\infty$, while the density of the Hamiltonian part in the asymptotic limit becomes  the  convolution of two semicircles. If actual finite-$N$ GUE or GOE eigenvalues  are used for the Hamiltonian spectrum, a small gap appears near $Y_H=0$, reflecting the level repulsion. The distributions are sampled for $N=100$, $R=10^4$, $\Gamma_D=1$, and $\alpha=0.5$, and averaged over $10^4$ independent realizations. 
}
\label{fig:ensembleA}
\end{figure*}

For real-valued jump operator eigenvalues, the dissipative contribution is real and nonpositive, $d_{ij}=-\frac12\sum_m\gamma_m(v_{mi}-v_{mj})^2$, while the Hamiltonian contribution is purely imaginary. In the large-$N$ limit, their independence implies a factorized spectral density,
\begin{equation}
 \displaystyle
 \rho_{\mathcal L}(x+iy)=p_D(x)\,\frac{1}{|\alpha|}\,
 \rho_{\Delta,2}\!\left(\frac{y}{\alpha}\right),
 \label{eq:Aproduct}
\end{equation}
where $\rho_{\Delta,2}=\rho_{\mathrm{sc},2}*\rho_{\mathrm{sc},2}$ is the standard self-convolution of the semicircle distribution (see Appendix \ref{app:convolution} for the exact expression).

In the limit $\alpha=0$, the imaginary marginal is $\delta(y)$. The density $p_D$ is the distribution of $-\frac12\sum_m\gamma_m(v_m-v_m')^2$ for independent drawn $v_m,v_m'$ (its characteristic function and moments are given in Appendix~\ref{app:ensembleA}). 

For identically distributed real jump eigenvalues with variance $\sigma_v^2$ and fourth central moment $\mu_4$, the dissipative marginal has mean $-\Gamma_D\sigma_v^2$ and variance $(\mu_4+\sigma_v^4)\sum_m\gamma_m^2/2$. In particular, with equal rates $\gamma_m=\Gamma_D/R$ and semicircular jump eigenvalues ($\sigma_v^2=1$, $\mu_4=2$), its large-$R$ limit is Gaussian, centered at $-\Gamma_D$ with standard deviation $\sigma_D=\Gamma_D\sqrt{3/(2R)}$.  Increasing parameter $\alpha$ stretches the imaginary marginal without compressing the real marginal. Figure~\ref{fig:ensembleA} illustrates this factorization.

For complex diagonal jump eigenvalues, the dissipative contributions to the eigenvalues read,
\begin{equation}
 d_{ij}=-\frac12\sum_m\gamma_m|v_{mi}-v_{mj}|^2
 +i\sum_m\gamma_m\operatorname{Im}(v_{mi}\overline{v_{mj}}).
 \label{eq:Acomplex}
\end{equation}
The dissipative component now also shifts the oscillation frequencies. If $\rho_D(x,y)$ denotes the joint dissipative spectral density, the full density is its convolution with the Hamiltonian difference and can be well approximated as (see Appendix~\ref{app:ensembleA})
{\small
\begin{equation}
\begin{aligned}
\rho_D(x+iy)
&=\frac{1}{2\pi\sqrt{\det\Sigma_D}}\\
&\hspace{-3em}
\exp\!\left[-\tfrac12
\tbinom{x-\mu_x}{y-\mu_y}^{T}
\!\Sigma_D^{-1}\!
\tbinom{x-\mu_x}{y-\mu_y}
\right],\,
\Sigma_D=
\begin{pmatrix}
\sigma_x^2 & \sigma_{xy}\\
\sigma_{xy} & \sigma_y^2
\end{pmatrix}.
\end{aligned}
\label{eq:rhoD_gaussian}
\end{equation}
}

The  Gaussian approximation to $\rho_D$ at large effective rank and the corresponding conditional-density formula are given in Appendix~\ref{app:ensembleA}.

The diagonal model considered in this section is fully analytically tractable. The spectra of Hamiltonian and dissipative components contribute to the  spectrum of the Lindbladian via the classical convolution along the imaginary axis. When both spectra are real, the interaction is absent, so we have the product of two distributions. 
  
\begin{figure*}[t]
\centering
\includegraphics[width=0.99\textwidth]{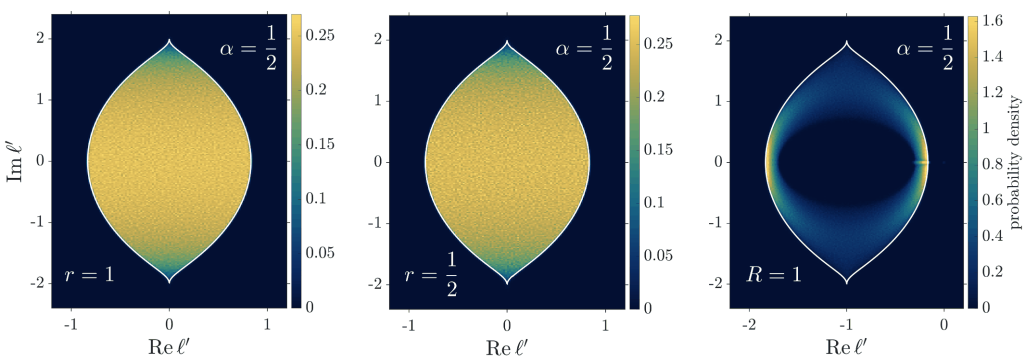}
\caption{Rescaled spectral densities, $\ell'=\sqrt{R}(\ell+1)$, of random channel-generated
Lindbladians (Ensemble B), Eqs.~(\ref{eq:lindblad3}) and (\ref{eq:lindblad4}),
for different channel ranks, for $N=100$ and
$\alpha=1/2$. From left to right: full rank
$r=R/(N^2-1)=1$, relative rank $r=1/2$, and
a single unitary channel, $R=1$.
The white lines show the lemon contour rotated
by $90^\circ$. The single-channel case exhibits a
strongly nonuniform spectral density while being confined to the lemon-shaped contour. Each distribution was sampled with $10^3$ realizations.}
\label{fig:ensembleB_ranks}
\end{figure*}

\subsection{Ensemble B: Channel-generated random Lindbladians}
\label{ensembleB}

Next we consider an ensemble of random Lindblad operators, Eq.~(\ref{eq:lindblad4}), with a specific dissipative part. Namely, when the latter is written in the form~\eqref{eq:lindblad3}, the CP map $\Psi$ is assumed to be trace preserving (TP). Hence $\Psidag(\oper)=\oper$, and the correction term reduces to the identity superoperator $\oper_{N^2}$. Therefore, the shifted operator $\cL_{\mathrm{D}}+\oper_{N^2}$ can be modeled, within the RMT framework, by a real Ginibre matrix,
$
\widehat{\cL}_{\mathrm{D}}+\oper_{N^2}\simeq  Y_R,
$
whose spectral radius is $1/\sqrt{R}$~\cite{bruzdaRandomQuantumOperations2009} (in the units of the rescaled spectrum $\ell'=N(\ell+1)$ introduced in Sec.~\ref{sec:lemon}). 

The Hamiltonian part,
$
\cL_H=-i\alpha[H,\cdot],
$
with $H$ being sampled from the GUE, has the same spectral properties as in the case of Ensemble A. Namely, its spectral density is given by the standard (classical) self-convolution of the semicircle, $\rho_{\Delta,2}=\rho_{\mathrm{sc},2}*\rho_{\mathrm{sc},2}$.
Note that, due to the symmetry of the semicircle at zero, the minus sign in $\mathbbm{1}\otimes H_0-\overline H_0\otimes\mathbbm{1}$ can be replaced with plus. 
Thus, at the RMT level, we obtain the same structure as in Eq.~\eqref{eq:model_1} but with the correction term  rescaled and rotated by $90^\circ$ in the complex plane, 
\begin{equation}
    \widehat{\mathcal{L}} = Y_R- 2i \alpha (\oper\otimes C + \overline C\otimes\oper),
\label{eq:ilemon}
\end{equation}
with $\operatorname{Tr}(Y Y^\dagger)=N^2$ and $
\operatorname{Tr}C^2=N/4$, as before.

Following the same path as for the lemon (see Appendix D in Ref.~\cite{denisovUniversalSpectraRandom2019}), we obtain the equation for the spectral contour
\begin{equation}
    \operatorname{Im}\left[i\alpha z+\GDl{2}\left(\frac{iz}{\alpha}\right)\right]=0,
    \label{eq:lemon_contour_modified}
\end{equation}
where $\GDl{2}$ is the Green's function of the density $\rD{2}$ (it is related to the function $\GDl{1}$ appearing in Eq.~\eqref{eq:lemon_contour} by $\GDl{2}(z)=\frac12 \GDl{1}(z/2)$).
In particular, for $\alpha=1/2$, this yields the lemon-shaped spectral support rotated by $90^\circ$; see the left panel of Fig.~\ref{fig:ensembleB_ranks}. The other panels show sampling results for random channel-generated Lindbladians, with the relative rank $r=R/(N^2-1)=1/2$ (middle) and for the total rank $R=1$ (right). The latter case corresponds to a unitary channel, for which the Lindbladian takes the form
$
\mathcal{L}_{U}(\rho)=-i[H,\rho]+U\rho U^\dagger-\rho
$.

The ultimate limit $R=1$ is interesting in itself. 
At the  RMT level, it corresponds to
$
\widehat{\mathcal{L}}_U=\overline U\otimes U-i2\alpha(\oper\otimes C
-\overline C\otimes\oper)
$
Remarkably, despite the different statistical structure of $\overline U\otimes U$ and a Ginibre matrix, the spectral boundary of the single-channel model agrees numerically with the prediction of Eq.~\eqref{eq:lemon_contour_modified}, although the spectral densities differ substantially.

\begin{figure*}[t]
\centering
\includegraphics[width=0.99\textwidth]{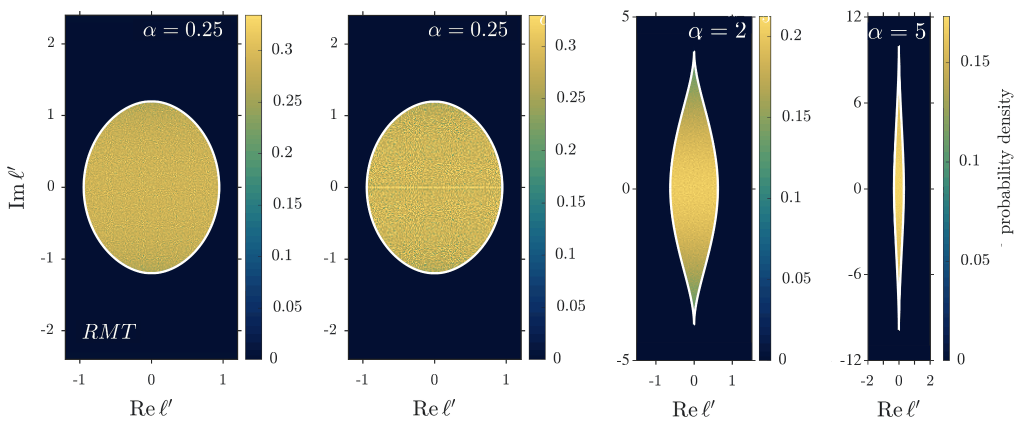}
\caption{Spectral densities of full-rank random
channel-generated Lindbladians (Ensemble B),
Eqs.~(\ref{eq:lindblad3}) and (\ref{eq:lindblad4}), and the corresponding
RMT model, Eq.~\eqref{eq:ilemon}, for different
values of the Hamiltonian weight $\alpha$.
From left to right: the RMT model and the
Lindbladian ensemble at $\alpha=0.25$, followed
by the Lindbladian ensemble at $\alpha=2$ and
$\alpha=5$. 
The white lines show the analytical spectral
boundaries given by
Eq.~\eqref{eq:lemon_contour_modified}.
Each spectral density was sampled over
$10^3$ realizations for $N=100$.}
\label{fig:ensembleB_alpha}
\end{figure*}

Figure~\ref{fig:ensembleB_alpha} presents the spectral densities of full-rank  Lindbladians sampled from Ensemble B for different values of $\alpha$.
As $\alpha$ increases, the spectral support becomes increasingly elongated along the imaginary axis, while remaining centered at $\operatorname{Re}\ell=-1$, independently of $\alpha$, since its position is determined solely by the strength of dissipation.

Notably, the model exhibits the following spectral property: In the limit $\alpha\to\infty$, the area of the spectral support approaches the  value $2\pi$ (in units where $Y_R$ has unit spectral radius). 

Following Feinberg and Zee~\cite{FeinbergNonHermitian1997},
we consider the random-matrix model
$\widehat{\mathcal L}=Y_R-i\alpha\mathcal D$, where
$\mathcal D=\oper\otimes H_0
-\overline{H_0}\otimes\oper$ is Hermitian and
$Y_R$ has unit spectral radius.
In the limit $N\to\infty$, the boundary of the spectral
support, $z=x+iy$, is determined by
\begin{equation}
\begin{aligned}
&\lim_{N\to\infty}\frac{1}{N^2}
\Tr\left[
(z\oper+i\alpha\mathcal D)
(z\oper+i\alpha\mathcal D)^\dagger
\right]^{-1}\\
&\qquad =
\int\frac{\rho_{\mathcal D}(k)\,\dd k}
{x^2+(y+\alpha k)^2}=1,
\end{aligned}
\end{equation}
where $\rho_{\mathcal D}$ is the normalized 
eigenvalue density of $\mathcal D$.

To evaluate the limit $\alpha\to\infty$, we substitute
$y=-\alpha u$ and $x=f(u)/\alpha$ into the boundary
condition. Using the representation of the Dirac delta
function,
$\lim_{\varepsilon\to0^+}
\varepsilon/(s^2+\varepsilon^2)=\pi\delta(s)$,
we obtain $f(u)=\pi\rho_{\mathcal D}(u)$ to leading order.
Thus, the spectral boundaries are given asymptotically by equation
$x_\pm(y)=\pm(\pi/\alpha)
\rho_{\mathcal D}(-y/\alpha)$.
Integrating the width of the spectral support along the
imaginary axis, we find
\begin{equation}
\begin{aligned}
\lim_{\alpha\to\infty}\mathcal A_{\mathrm{CPTP}}
&=\lim_{\alpha\to\infty}
\int [x_+(y)-x_-(y)]\,\dd y\\
&=2\pi\int\rho_{\mathcal D}(u)\,\dd u
=2\pi.
\end{aligned}
\label{eq:CPTP_2pi}
\end{equation}
Hence, as $\alpha\to\infty$, the spectral support
stretches along the imaginary axis while its width
in the real direction decreases as $\alpha^{-1}$.
These effects compensate, and the total area enclosed by the spectral contour approaches the universal value $2\pi$. 
We return to this \textit{spectral squeezing} phenomenon in
Section~\ref{sec:squeezing}.

\subsection{Ensemble C: Generic random Lindbladians}
\label{ensembleC}

In the third ensemble the Hamiltonian and dissipative components of
Eq.~\eqref{eq:lindblad4} are sampled independently, without the commutativity of Ensemble A or the trace-preserving restriction on the
seed CP map of Ensemble B. The main difficulty is already visible after
vectorization, Eq.~(\ref{eq:RMT_general}). Namely, each of the Hermitian corrections acts on one of the two
tensor factors, while the Hamiltonian contribution has the same structure
but is anti-Hermitian. Terms acting on different factors commute so  they
are therefore not freely independent. Consequently, the eigenvalue law of
their tensor sum is a classical convolution, and free additive convolution cannot be used to combine its two legs.

The distinction is not minor. For example, at $\alpha=0$
let $\mathsf A=C_A\otimes\oper/2$ and
$\mathsf B=\oper\otimes C_B/2$, with $C_A$ and $C_B$ being independent normalized
GUE matrices. Then the alternating mixed moment is
$\varphi(\mathsf A\mathsf B\mathsf A\mathsf B)
=\varphi(\mathsf A^2)\varphi(\mathsf B^2)=1/16$, whereas it
would vanish for centered free variables. The fourth moment of
$\mathsf A+\mathsf B$ is $5/8$, rather than the value $1/2$
obtained by a free-convolution approximation. 



As in Sec.~\ref{sec:lemon}, the real Ginibre
matrix can be replaced by a complex Ginibre one if we plan to derive the asymptotic
spectral boundary. We therefore consider the matrix with the following structure, 
\begin{equation}
\begin{aligned}
X=Y+W_A\otimes\oper_B+
\oper_A\otimes W_B = Y + D_{\alpha},
\end{aligned}
 \label{Lmodel-1}
\end{equation}
where $W_A=\frac{1}{2}C_A+i\alpha H_A$ and $
W_B=\frac{1}{2}C_B+i\alpha H_B$
are two independently sampled elliptic matrices~\cite{Jarosz2006JPhysA},
and $Y$ is a complex Ginibre matrix with entry variance
$v_Y/(N_AN_B)$. The overall sign and conjugation in Eq.~\eqref{eq:RMT_general} are absorbed into the independent Gaussian copies $W_A$ and $W_B$, which have the same elliptic law as $W$. 

The two independent copies in $D_{\alpha}$, Eq.~\eqref{Lmodel-1}, replace the two
conjugate tensor legs of the physical construction. Their normalized
fixed-order tensor trace moments are the same in the large-$N$ limit. The assumption of independence, however,  helps us to evaluate the spectral boundary analytically. 

Consider now matrix $W$ belonging to the Gaussian elliptic ensemble~
\cite{Jarosz2006JPhysA}. Its two covariance parameters are
$\sigma^2=\tfrac14+\alpha^2$ and
$c\equiv\tau\sigma^2=\tfrac14-\alpha^2$, so that
$\mathbb E[W_{ij}\overline W_{kl}]=
(\sigma^2/N)\delta_{ik}\delta_{jl}$ and
$\mathbb E[W_{ij}W_{kl}]=(c/N)\delta_{il}\delta_{jk}$.
The limiting support of $W$ is an ellipse with semiaxes
$(\sigma^2+c)/\sigma=1/(2\sigma)$ and
$(\sigma^2-c)/\sigma=2\alpha^2/\sigma$. Thus the tensor term
$D=W_A\otimes\mathbbm1_B+\mathbbm1_A\otimes W_B$ has an
elliptic eigenvalue support with twice these semiaxes. Yet again, we emphasize that this planar
eigenvalue density alone is insufficient for the addition of $Y$ because 
$D_{\alpha}$ is non-normal for $\alpha>0$ and its resolvent depends on
information beyond the one contained in the eigenvalues. We could retain this information through an operator-valued resolvent~\cite{HeltonRashidiFarSpeicher2007} rather than applying the scalar construction used in  Sec.~\ref{sec:lemon}.

\subsection{Hermitization, matrix Dyson equation, and the boundary integral}
\label{sec:ensembleC_boundary}

For a matrix $X$, Eq.~\eqref{Lmodel-1}, the ordinary resolvent is
$R_X(z)=(z\mathbbm1-X)^{-1}$, defined for
$z\notin\operatorname{spec}X$. Its trace is analytic outside
the spectrum, but the eigenvalue density of a non-Hermitian matrix
cannot be obtained by evaluating this analytic resolvent inside the
spectrum. We therefore introduce the Hermitian matrix
\begin{equation}
 \mathcal H_z(X)=
 \begin{pmatrix}
  0&z\mathbbm1-X\\
  \overline z\mathbbm1-X^\dagger&0
 \end{pmatrix},
 \label{eq:ensembleC_hermitization}
\end{equation}
and use its resolvent at the imaginary regulator $w>0$.
Equivalently, after exchanging block columns, we work with the
regularized hermitized Green function~\cite{FeinbergNonHermitian1997}
\begin{equation}
\mathsf R_X(z,w)=
\begin{pmatrix}
 z\mathbbm1-X&iw\mathbbm1\\
 iw\mathbbm1&\overline z\mathbbm1-X^\dagger
\end{pmatrix}^{-1}.
\label{eq:ensembleC_resolvent_def}
\end{equation}
The regulator makes this inverse well defined and the limit
$w\to0^+$ should be taken after the large-matrix limit. The off-diagonal
blocks distinguish the exterior analytic solution from the
non-holomorphic solution carrying the spectral density.

The key simplification is to average over the random matrices in two
steps. We first fix one realization of $W_A$ and average over $Y$
and $W_B$, taking the normalized trace only over the $B$ factor.
The result remains a $2\times2$ block matrix
$\mathsf G_A=(G_{ab})_{a,b=1}^2$, whose blocks act on
$\mathbb C^{N_A}$. Only after solving for these blocks we average $N_A^{-1}\Tr_A$ over $W_A$. We denote the resulting expectation by
\begin{equation}
\varphi(T)=\lim_{N_A\to\infty}
 \mathbb E_{W_A}\frac1{N_A}\Tr_A T.
 \label{eq:ensembleC_normalized_trace}
\end{equation}
We call it \emph{quenched-$A$} procedure in which the $A$ matrix is held
fixed during the first averaging step, not replaced by a scalar.

The first step also defines the conditional, operator-valued
Cauchy transform
\begin{equation}
\mathsf G_A(z,w;W_A)
=
\lim_{N_B\to\infty}
\mathbb E_{Y,W_B}
\left[
(\mathrm{id}_A\otimes\mathrm{tr}_B)
\mathsf R_X(z,w)
\right], \nonumber
\end{equation}
where $\mathrm{tr}_B=N_B^{-1}\mathrm{Tr}_B$ acts
on each block of the hermitized resolvent.
Thus, $\mathsf G_A$ is a $2\times2$ block matrix
whose entries are $N_A\times N_A$ matrices.
In particular, the dependence on $W_A$ is
retained rather than replaced by its ensemble
average. This construction is an instance of
the operator-valued Cauchy-transform formalism
\cite{HeltonRashidiFarSpeicher2007}.

In the large-$N_B$ limit, the conditional
Cauchy transform satisfies the following
self-consistent matrix Dyson equation (MDE):
\begin{equation}
\begin{aligned}
\mathsf G_A^{-1}&=\mathsf Q_A-\mathsf W_A-\mathcal S[\mathsf G_A],\\
\mathsf Q_A&=\begin{pmatrix}z\oper_A&iw\oper_A\\
 iw\oper_A&\overline z\oper_A\end{pmatrix},~
\mathsf W_A=\operatorname{diag}(W_A,W_A^\dagger),
\end{aligned}
\label{eq:ensembleC_MDE_definition}
\end{equation}
where $q_{ab}=N_A^{-1}\Tr_A G_{ab}$ and the self-energy map is
\begin{equation}
\mathcal S[\mathsf G_A]=
\begin{pmatrix}
 cG_{11}&\sigma^2G_{12}\\
 \sigma^2G_{21}&cG_{22}
\end{pmatrix}
+v_Y\begin{pmatrix}
 0&q_{12}\mathbbm1_A\\
 q_{21}\mathbbm1_A&0
\end{pmatrix}.
\label{eq:ensembleC_MDE_selfenergy}
\end{equation}
Thus the MDE is an explicit nonlinear equation for a
$2N_A\times2N_A$ matrix, is not a scalar-valued Dyson equation obtained by incorrectly assuming freeness of the two tensor factors~
\cite{HeltonRashidiFarSpeicher2007,Speicher2011,MingoSpeicher2017}.
It's derivation and structure  are discussed in details in Appendix~\ref{sec:appendix_A_Quenched}.

Outside the spectral support, taking $w\to0^+$ gives
$G_{12}=G_{21}=0$. The upper diagonal block $M=G_{11}$ thus satisfies
\begin{equation}
M=[z\mathbbm1_A-W_A-cM]^{-1},\qquad
M(z)\sim z^{-1}\mathbbm1_A
\label{eq:ensembleC_holomorphic_def}
\end{equation}
as $|z|\to\infty$ so that the asymptotics select the physical exterior
branch. This solution ceases to describe a stable exterior
resolvent when an arbitrarily small off-diagonal perturbation is
amplified by the Dyson equation. Linearizing the off-diagonal block
gives the map $\Lambda_z[T]=\sigma^2MTM^\dagger$ and the regular
spectral-boundary condition
\begin{equation}
\begin{aligned}
 v_Y\Phi(z)=1,~
 \Phi(z)=\sum_{n\ge1}(\sigma^2)^{n-1}
\varphi\big[M^n(M^\dagger)^n\big].
\end{aligned}
\label{eq:ensembleC_phi_def}
\end{equation}
The series is non-negative and represents the trace of the
exterior resolvent product of the source $D$. It is evaluated only
where the holomorphic solution and the associated Neumann expansion
are admissible (see Appendix~\ref{sec:appendix_A_Quenched} for details).

To eliminate the remaining matrix average, we use the exterior
Stieltjes transform of the elliptic matrix $W$,
$g(\lambda)^{-1}=\lambda-cg(\lambda)$,
$g(\lambda)\sim\lambda^{-1}$, and the exterior algebraic
function $f_z(\lambda)^{-1}=z-\lambda-cf_z(\lambda)$,
$f_z(\lambda)\sim(z-\lambda)^{-1}$. Then
$M=f_z(W_A)$, wherever the analytic functional calculus applies.
For a contour $\Gamma$ enclosing the elliptic support of $W_A$,
we obtain the integral equation
\begin{widetext}
\begin{equation}
\begin{aligned}
\Phi(z)&=\sum_{n=1}^{\infty}(\sigma^2)^{n-1}
 \sum_{k=0}^{\infty}(\sigma^2)^k|a_{nk}(z)|^2
 =\frac1{v_Y},~~
a_{nk}(z)&=\oint_\Gamma\frac{\dd\lambda}{2\pi i}\,
 f_z(\lambda)^n g(\lambda)^{k+1}.
\end{aligned}
\label{eq:ensembleC_integral_boundary}
\end{equation}
\end{widetext}
Equations~\eqref{eq:ensembleC_phi_def} and \eqref{eq:ensembleC_integral_boundary} are our central analytic result. Note that they involves only one-dimensional
contour integrals of explicitly known algebraic functions and does not require sampling 
of some specific $N\times N$ matrices. For comparison, the equivalent
double-contour integral obtained by resumming the geometric series is presented in
Appendix~\ref{sec:appendix_B_boundary}. We discuss the numerical treatment of Eqs.~(\ref{eq:ensembleC_integral_boundary}, \ref{eq:ensembleC_integral_boundary}) in Section~\ref{sec:ensembleC_numerics}.


\begin{figure*}[!t]
 \centering
 \includegraphics[width=0.97\linewidth]{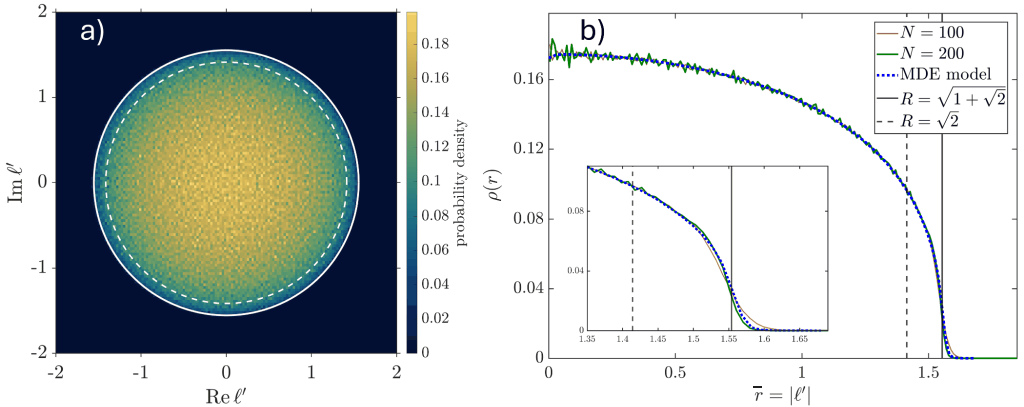}
 \caption{Spectral properties of generic random Lindbladians
(Ensemble C) at the balanced point $\alpha=1/2$.
(a) Spectral density of the rescaled Lindbladian
$\ell'=N(\ell+1)$ for $N=100$, sanpled with
$10^3$  realizations. The solid white circle
shows the analytical spectral boundary of radius
$\bar r=\sqrt{1+\sqrt{2}}$, Eq.~\eqref{eq:ensembleC_silver_radius},
while the dashed circle indicates the radius
$\bar r=\sqrt{2}$ predicted by the elliptic Gaussian approximation.
(b) Radial spectral density $\rho(\bar r)$, Eq.~\eqref{eq:radial}, obtained
by numerical sampling for $N=100$ and $N=200$
and by solving the regularized operator-valued matrix
Dyson equation, Eq.~\eqref{eq:ensembleC_MDE_definition}.
The inset magnifies the region near the spectral edge.
}
 \label{fig:ensembleC_radial}
\end{figure*}

\subsection{Two exactly solvable limits: $\alpha=0$ and $\alpha=1/2$}
\label{sec:ensembleC_silver}

The first test is the purely dissipative limit $\alpha=0$. In this case $W_A$ and $W_B$ are Hermitian. The tensor
spectrum is the classical self-convolution of two semicircles,
and the stability function reduces to
$\Phi(z)=-\Im G_{\Delta,1}(z)/\Im z$. Thus, for the physical normalization $v_Y=1$,
$v_Y\Phi=1$ becomes
$\Im[z+G_{\Delta,1}(z)]=0$, reproducing the lemon contour
of Eq.~\eqref{eq:lemon_contour}~\cite{denisovUniversalSpectraRandom2019}.
The detailed reduction is presented in Appendix~\ref{sec:appendix_C_lemon}.

At the balanced point $\alpha=1/2$, $c=0$, $\sigma^2=1/2$,
and the support becomes a disk. With $\zeta=|z|^2$ one finds
$\Phi(\zeta)=[\zeta(\zeta-4\sigma^2)]^{-1/2}$ (see Appendix~\ref{sec:appendix_D_silver}), so
\begin{equation}
 \bar r=\sqrt{1+\sqrt2}\simeq1.553774.
 \label{eq:ensembleC_silver_radius}
\end{equation}
We refer to this value as the \emph{silver radius}, since
$r^2=1+\sqrt2=[2;2,2,\dots]$ is the silver ratio
\cite{Spinadel1999}.  In particular, the support is not a disk of radius 
$\sqrt2$ as predicted by the second-moment-matched elliptic Gaussian approximation; see Fig.~\ref{fig:ensembleC_radial}.


\begin{figure*}[t]
 \centering
 \includegraphics[width=0.97\linewidth]{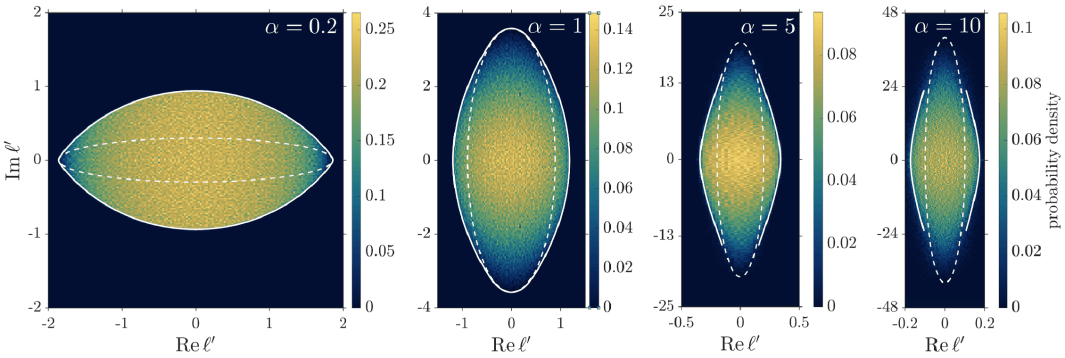}
 \caption{Ensemble C: Spectral densities of rescaled random
 Lindbladians, $\cL_{\alpha}'=N(\cL_{\alpha}+\oper_{N^2})$, for $\alpha=0.2$, $1$, $5$, and
 $10$ (left to right). Solid white curves indicate the spectral boundary obtained for the RMT model $X=Y+D_{\alpha}$  
 by solving  Eqs.~\eqref{eq:ensembleC_phi_def} and \eqref{eq:ensembleC_integral_boundary}.
 Dashed white ellipses mark the eigenvalue support of the tensor
 term $D_{\alpha}$ alone.
 For $\alpha=5$ and $10$ the solid curves terminate where
 convergence or numerical validation of the numerical solution
 becomes unreliable. The colored histograms show the planar eigenvalue density obtained by direct sampling of random Lindbladians with $N=100$, using $10^3$ realizations for each value of $\alpha$.
Note that the axis ranges and color scales differ between panels.}
 \label{fig:ensembleC_generalalpha}
\end{figure*}

\subsection{Numerical evaluation of the boundary integral}
\label{sec:ensembleC_numerics}

For arbitrary $\alpha$, the two algebraic functions $f_z$ and $g$
are known explicitly, but the positive double series in
Eq.~\eqref{eq:ensembleC_integral_boundary} has to be evaluated
numerically. We calculate the integrals directly  by parameterizing
the integration path in terms of the exterior Stieltjes
transform $g(\lambda)$, introduced above. Its defining
equation can be inverted to give
$\lambda=g^{-1}+cg$.
We choose the values of $g$ on a circle of radius
$q/\sigma$, with $0<q<1$,
\begin{equation}
g(\lambda_q(\theta))
=\frac{q}{\sigma}e^{-i\theta},
~
\lambda_q(\theta)
=\frac{\sigma}{q}e^{i\theta}
+\frac{cq}{\sigma}e^{-i\theta},
~ 0\leq\theta<2\pi. \nonumber
\end{equation}
The second expression follows directly by substituting
the first into $\lambda=g^{-1}+cg$.
The parameterization describes an ellipse enclosing
the spectral support of $W_A$. At $q=1$, it reduces
to the boundary of that support, whereas $q<1$
places the integration contour outside it.
The value of $q$ must be chosen so that the exterior
branch of $f_z(\lambda)$ remains analytic on and
inside the integration contour. For $q$ sufficiently close to one this ellipse
surrounds the spectral support of $W_A$, while remaining outside
it. The chosen contour must also lie in a region where the selected
branch of $f_z$ is analytic; this second requirement becomes
restrictive near the unresolved tips at large $\alpha$.

For each trial value $z$, we evaluate
$f_z(\lambda_q(\theta))$ on an equally spaced angular grid.
The quadratic equation for $f_z$ has two roots. We select the
solution asymptotic to $(z-\lambda)^{-1}$ and follow that root
continuously around the ellipse, thereby avoiding artificial
switches between quadratic branches. If
\begin{equation}
 F_{nk}(z)=\frac1{2\pi}\int_0^{2\pi}
 f_z(\lambda_q(\theta))^n e^{-ik\theta}\,\dd\theta,
 \label{eq:ensembleC_fft_coeff}
\end{equation}
then the coefficients entering the contour equation satisfy
\begin{equation}
 a_{nk}(z)=\left(\frac q\sigma\right)^k
 \left[F_{nk}(z)-\frac{cq^2}{\sigma^2}F_{n,k+2}(z)\right].
 \label{eq:ensembleC_fft_an}
\end{equation}
Consequently, the quantity required by the boundary search is
\begin{equation}
\begin{aligned}
 \Phi(z)=\sum_{n\ge1}(\sigma^2)^{n-1}
 \sum_{k\ge0}q^{2k}\Big|F_{nk}(z)
 -\frac{cq^2}{\sigma^2}F_{n,k+2}(z)\Big|^2.
\end{aligned}
\label{eq:ensembleC_Fourier_main}
\end{equation}
The Fourier coefficients are computed simultaneously by a fast
Fourier transform (FFT). This representation is preferable to
resumming the entire series under a double contour integral because
its terms are non-negative, and no cancellation between large
oscillatory contributions is needed. The derivation leading to
Eqs.~\eqref{eq:ensembleC_fft_an} and
\eqref{eq:ensembleC_Fourier_main} is given in
Appendix~\ref{sec:appendix_B_boundary}.

The order-$n$ contribution is calculated recursively using
$h_1=f_z$ and $h_{n+1}=(\sigma f_z)h_n$ on the angular grid.
After computing the FFT of each $h_n$, we sum its positive
contributions over the Fourier index $k$. We increase the number
of outer terms until the contribution of several consecutive
blocks is small compared with the accumulated $\Phi(z)$.
In practice, monitoring a whole positive block is more reliable
than stopping after a few accidentally small individual terms.
The largest Fourier index is also controlled by the angular
grid, which must be refined independently of the $n$ cutoff.

We locate the boundary along a set of polar directions. For each
direction, the search starts at a distant exterior point with
$\Phi(z)<1/v_Y$, moves inward, and brackets the root
$\Phi(z)-1/v_Y=0$ when the intermediate evaluations remain
admissible. The previous angle supplies a continuation guess,
while the reflection symmetries of the ensemble allow us to
calculate the first quadrant and obtain the others by reflection.
Each accepted root is independently checked by moving the Cauchy
contour from $q_1$ to $q_2>q_1$, by doubling the angular grid,
and by verifying both the positive series tail and the residual
$|v_Y\Phi(z)-1|$. A failed or divergent evaluation is marked
'unresolved'.
The numerical tolerances, convergence tests,
and root-search logic are detailed in
Appendix~\ref{app:EC_numerics}.

An alternative numerical method solves the finite-$N_A$ matrix
Dyson equation on its exterior branch and sums the stability
series through the \emph{Stein equation}
\begin{equation}
 Z=\oper_A+\sigma^2MZM^\dagger,~
 \Phi(z)=\frac{\varphi(Z)-1}{\sigma^2}.
 \label{eq:ensembleC_Stein_main}
\end{equation}
This auxiliary equation is numerically distinct from the direct
FFT integral and provides a useful consistency check. Its
Neumann-series interpretation requires
\begin{equation}
 \operatorname{spr}(\Lambda_z)
 =\sigma^2\operatorname{spr}(M)^2<1,
 ~\Lambda_z[T]=\sigma^2MTM^\dagger.
 \label{eq:ensembleC_Stein_condition_main}
\end{equation}
We refer to the loss of this inequality as the \emph{Stein
convergence threshold}. It is a threshold for the exterior
Neumann expansion, not the condition $v_Y\Phi=1$ for the
spectral boundary of $X$. For example, at $\alpha=1/2$ the
threshold occurs at $|z|=\sqrt2$, whereas the physical boundary
lies at $|z|=\sqrt{1+\sqrt2}$. Even where an algebraic solution
of the Stein equation exists, it cannot automatically be used
outside the convergent regime. The full planar density requires
the regularized, non-holomorphic MDE of
Eq.~\eqref{eq:ensembleC_MDE_definition}, not merely the exterior
boundary equation (Appendix~\ref{sec:appendix_A_Quenched}).


The balanced case $\alpha=1/2$ provides a
first numerical check independent of the contour integration.
Panel (a) of Fig.~\ref{fig:ensembleC_radial} shows the distribution sampled with  $10^3$ realizations for $N=100$. The observed edge follows the solid
circle of the silver radius
$\bar r=\sqrt{1+\sqrt2}\simeq1.554$, which is in full agreement with theory, and again disclosing smaller dashed
circle at $\bar r=\sqrt2$ obtained from the elliptic Gaussian approximation.
Panel (b) compares the
radial planar density 
\begin{equation}
\begin{aligned}
\rho(\bar r)=\frac{1}{2\pi \bar r}\frac{dF(\bar r)}{d\bar r},
~
F(\bar r)=\int_{|z|\le\bar r}\rho(z)\,d^2z,
\end{aligned}
\label{eq:radial}
\end{equation}
obtained by direct sampling for $N=100$
and $N=200$ with the numerical solution of the regularized,
non-holomorphic MDE. The latter is obtained for the
regulator value $w= 10^{-3}$ (further details are given in Appendix~\ref{app:quenchedA_convergence}). 
The good agreement confirms the validity of the hermitized MDE at the
level of the density, not only the spectral support.

Figure~\ref{fig:ensembleC_generalalpha} compares the rescaled spectral 
densities $\ell'=N(\ell+1)$ of genuine random Lindbladians with the spectral support predicted by the tensor model for $\alpha=0.2,1,5$, and $10$.  The solid white curves are
portions of the boundary of the full model $X=Y+D_{\alpha}$ obtained
by solving $v_Y\Phi=1$ along the regular exterior branch.
The dashed white ellipses  show
the  spectral support of the tensor term $D_{\alpha}$ alone,
with the semiaxes specified in Sec.~\ref{ensembleC}. 

For $\alpha=0.2$, the density is horizontally extended and still close
to the lemon-shaped  support. The ellipse for 
$D_{\alpha}$ spreads under the addition of $Y$  over a substantially
larger range of imaginary parts.  Increasing the weight to $\alpha=1$
makes the spectral cloud  more elongated vertically.
The ellipse corresponding to the spectral contour of $D_{\alpha}$ remains inside the support of the full sum.

Further increase to $\alpha=5$ and $10$ results in a pronounced squeezing
in the real direction combined with a large stretch in the imaginary direction. 
Near the upper and lower tips at large $\alpha$, the solid curves
terminate before forming closed contours. This termination
is not a measured endpoint of the eigenvalue support. The direct
integral method uses the exterior resolvent of $D_{\alpha}$; when the
candidate boundary approaches $\operatorname{spec}(D_{\alpha})$, its
analytic domain becomes restrictive and the positive series can
converge extremely slowly or fail to converge. At the same time,
the integration contour must remain outside the elliptic support
without crossing singularities of $f_z$, which reduces the
numerically useful range of $q$. Refining the FFT grid does not
necessarily cure an instability caused by contour dependence or
a genuinely inadmissible exterior branch.

The algorithm therefore discards an angular direction
if the series diverges, its positive tail is unresolved, or the
$q$-deformation and quadrature checks cannot both be satisfied.
In particular, failure of a trial point with $\Phi<1/v_Y$ is
not counted as an interior crossing. Hence the last plotted
point is the last numerically accepted boundary
point. 


There is an interesting empirical relation between the loss of stability of the numerical procedure and the support criterion for adding free circular noise to a non-normal operator discussed in Refs.~\cite{zhong2025brownmeasuresumelliptic,erdos2024densitybrownmeasurefree}. In these works, sharp and quadratic types of spectral density cutoff are established under an assumption that places the spectrum of the unperturbed operator (that is $\operatorname{spec}(D_{\alpha})$ in our case) strictly inside the support of the circularly perturbed operator. As can be seen from the results for $\alpha=5$ and $10$, the loss of numerical stability occurs close to the point where this condition appears to cease to hold. At the same time, the spectral density near the tips becomes  more diffuse, and the sharpness of the boundary is visibly reduced. It would be interesting to understand whether this correspondence is merely coincidental or reflects a deeper connection between the stability of the operator-valued MDE and the regularity conditions controlling the Brown-measure density near the boundary.


\section{Spectral squeezing}
\label{sec:squeezing}

The three ensembles discussed above respond  to an increase of the Hamiltonian weight $\alpha$ in qualitatively different ways.

We refer to the following behaviour as \emph{spectral squeezing}: In the thermodynamic limit, the geometric area of the spectral support approaches a finite, ensemble-dependent constant as $\alpha\to\infty$. The name also refers to the lemon-shaped support at the limit $\alpha=0$. Namely, as if squeezing a lemon, increasing $\alpha$ stretches the spectral bulk in one direction and compresses it in the other, without unbounded growth of its area; see Fig.~\ref{fig:fig0}.

There is no squeezing effect for Ensemble A. The real and imaginary parts of the spectrum are statistically independent for real jump eigenvalues, and the spectral density is the product density of Eq.~\eqref{eq:Aproduct}. Thus, increasing the weight $\alpha$ simply stretches the distribution along the imaginary axis while leaving the real marginal unchanged, so that there is no compensating compression in the real direction. For complex jump eigenvalues, the convolution in the imaginary direction couples the real and imaginary parts of the eigenvalue distribution. Nevertheless, varying $\alpha$ still leaves the real marginal unchanged and only broadens the distribution in the imaginary direction. Thus, also in this case, no spectral squeezing occurs.

For Ensemble B, in contrast, there is a squeezing effect. The area enclosed by the spectral contour grows monotonically from $\pi$ (the unit disk at $\alpha=0$, in units where $Y_R$ has unit spectral radius) and saturates at $2\pi$ as $\alpha\to\infty$, see Eq.~\eqref{eq:CPTP_2pi}.

For Ensemble C, the squeezing mechanism is already visible in the tensor term $D_\alpha$. Its spectral support has semiaxes $a_x=1/\sigma$ and $a_y=4\alpha^2/\sigma$, with $\sigma^2=\alpha^2+1/4$, and therefore $a_x\sim\alpha^{-1}$ and $a_y\sim4\alpha$ as $\alpha\to\infty$. Consequently, the area of the support of $D_\alpha$ approaches $4\pi$. The addition of the circular component $Y$ modifies this boundary according to the quenched-$A$ condition, Eq.~\eqref{eq:ensembleC_phi_def}. The numerical results indicate that the full spectral support retains the same qualitative large-$\alpha$ scaling, being stretched vertically and compressed horizontally, and therefore suggest that its area approaches a finite nonzero limit. The incomplete contours near the upper and lower tips, however, do not allow us to determine this limiting area at present.

This behavior, established for Ensemble B and conjectured for Ensemble C, corresponds to an asymptotically area-preserving deformation of the spectral bulk.  The support is elongated along the imaginary axis, $y_{\max}\sim\alpha$, and compressed along the real axis, $x_{\max}\sim\alpha^{-1}$, while their product stays finite. This is geometrically reminiscent of squeezed Gaussian states in quantum optics, where the quadrature variances are rescaled in opposite directions while the area of the covariance ellipse is preserved~\cite{Yuen1976two-photon}. Here the role of phase space is played by the spectral plane. Whether the geometric area of the spectral support of a Lindbladian admits a direct operational interpretation or is an implicitly measurable physical quantity,  remains an interesting open question.

\section{Concluding remarks}
\label{sec:conclusion}


From a mathematical perspective, random Lindbladians constitute an interesting class of non-Hermitian random matrices~\cite{andersonGuionnetZeitouni2010}.  To date, the only analytical results concerning the geometric structure
of the spectra of random Lindbladians were obtained for purely
dissipative ensembles, whose characteristic lemon-shaped spectral support
was derived using free-probability methods~\cite{denisovUniversalSpectraRandom2019,Tarnowski2021}.


In this work, we go beyond the pure dissipative limit and study the spectral properties of random Lindblad generators for an arbitrary relative weight of the Hamiltonian and
dissipative components. Standard methods based on the Blue function~\cite{Jarosz2006JPhysA} and the scalar second-moment approximation are insufficient because of the tensor-product structure of the problem. We therefore develop an approach
based on operator-valued free probability~
\cite{HeltonRashidiFarSpeicher2007,AjankiErdosKruger2019}.
By averaging over one tensor factor while keeping the randomness in the other fixed, we obtain a matrix Dyson equation. Its analysis leads to an explicit equation for
the spectral boundary, expressed through one-dimensional contour integrals of known algebraic functions. We evaluate these integrals numerically, using a Fourier representation, that allowed  us to determine the spectral
boundary for different Hamiltonian strengths.

At the balanced point $\alpha=1/2$, where the spectral support becomes rotationally symmetric, we obtain an exact analytical expression for its radius and independently
calculate the radial spectral density by solving the regularized matrix Dyson equation. Direct numerical sampling of random Lindbladians confirms the RMT predictions.


An interesting question is how locality in the Hamiltonian and dissipative parts affects the spectral properties of random Lindbladians. Wang, Piazza, and Luitz~\cite{luitz2020PRL} demonstrated that restricting jump operators to act on a limited number of spins splits the spectrum of a purely dissipative random Lindbladian into distinct clusters associated with different relaxation timescales. These clusters merge into the familiar lemon-shaped support when the locality restriction disappears. It would be interesting to determine how this scenario changes when local dissipation is combined with a fully nonlocal random Hamiltonian or, conversely, when a local Hamiltonian is coupled to nonlocal dissipation. A related setting is provided by the dissipative Sachdev--Ye--Kitaev (SYK) model~\cite{sa2022lindbladian,kulkarni2022lindbladian,kawabata2023PRB}, which can combine random few-body jump operators with an all-to-all interacting Hamiltonian.

Another interesting question is how symmetries affect the spectral properties of random Lindbladians. Recent classifications of Lindbladians assign them to distinct non-Hermitian symmetry classes with characteristic spectral properties~\cite{sa2023PRX,kawabata2023PRXQ}. Adding a Hamiltonian component may preserve or break symmetries imposed by the dissipative part, depending on the properties of both. It would be useful to quantify such symmetry breaking through spectral observables and determine, e.g., how the spectral support varies between different symmetry classes of random Lindbladians.

Correlations between complex eigenvalues~\cite{saComplexSpacingRatios2020a,Tarnowski2021} provide the next level of spectral characterization of random Lindbladians, allowing us to explore universality beyond the spectral density and the geometry of its support. For chaotic dissipative systems, these correlations are well described by Ginibre random-matrix statistics~\cite{akemannUniversalSignatureIntegrability2019,saComplexSpacingRatios2020a}. It is an interesting task to determine how they change as the Hamiltonian contribution becomes dominant, particularly when the Hamiltonian is integrable rather than chaotic. An integrable Hamiltonian combined with generic dissipation need not produce an integrable Lindbladian, and identifying the conditions under which its spectral correlations retain signatures of integrability is an interesting problem, especially in the context of the emerging theory of dissipative quantum chaos~\cite{sa2026whatwetalk}.

Finally, the quantum-to-classical transition induced by superdecoherence~\cite{Tarnowski2021} raises another question concerning the role of the Hamiltonian component. Superdecoherence suppresses the coherent components of the Choi operator corresponding to the given Lindbladian and, in the limit of complete superdecoherence, produces a classical Kolmogorov generator~\cite{Liggett2010}. The Hamiltonian contribution vanishes in this limit (see the discussion in the conclusion of Ref.~\cite{Tarnowski2021}), so, for a fixed decoherence basis, Lindbladians with the same dissipative part but different Hamiltonians have the same 'classical shadow'. However, the superdecoherence-induced  evolution~\cite{Tarnowski2026ManyFacets} towards the same Kolmogorov operator may depend substantially on the Hamiltonian component. Understanding how different Hamiltonians affect this transition, including a possible emergence of intermediate spectral structures, would reveal new aspects of the interplay between coherent and dissipative dynamics.


\section*{Acknowledgments}
We are grateful to D. Chru{\'s}ci{\'n}ski and  M. 
A.~Nowak for useful discussions. MH acknowledges that the study was funded by “Research support module” as part of the “Excellence Initiative - Research University” program at the Jagiellonian University in Kraków. MH and KZ  acknowledge financial support from the European Union through the ERC Advanced Grant TAtypic, Project No. 101142236. SD acknowledges support from the Norwegian Research Council through the Norwegian Quantum Software Center (NorQSoft, project number 361350). WT, MH, and KZ thank the Department of Computer Science at Oslo Metropolitan University for its hospitality during research visits that contributed to the development of this work. 


\appendix
\section{Spectral distributions for Lindblad operators with commuting Hamiltonian and jump operators}
\label{app:ensembleA}
\subsection{Real diagonal jumps: exact convolution and moments}
Let $v,v'$ be independent copies of a real variable with mean $\mu_v$, variance $\sigma_v^2$ and fourth central moment $\mu_4$. Set $\theta=(v-v')^2/2$. For independent channels, $d=-\sum_m\gamma_m\theta_m$. With $\varphi_\theta(t)=\mathbb E e^{it\theta}$,
\begin{equation} \nonumber
 \varphi_D(t)=\prod_{m=1}^R\varphi_\theta(-\gamma_m t),~~
 \mathbb E\theta=\sigma_v^2,~~
 \operatorname{Var}(\theta)=\frac{\mu_4+\sigma_v^4}{2}.
\end{equation}
It follows that
\begin{equation} \nonumber
 \mathbb Ed=-\Gamma_D\sigma_v^2,\qquad
 \operatorname{Var}(d)=\frac{\mu_4+\sigma_v^4}{2}\sum_m\gamma_m^2.
\end{equation}
For positive rates define $R_{\mathrm{eff}}=\Gamma_D^2/\sum_m\gamma_m^2$. Then $\operatorname{Var}(d)=\Gamma_D^2(\mu_4+\sigma_v^4)/(2R_{\mathrm{eff}})$. For equal rates, $R_{\mathrm{eff}}=R$. Under the usual Lindeberg condition on the weighted summands, $(d-\mathbb Ed)/\sqrt{\operatorname{Var}(d)}$ converges to a standard normal variable. The Wigner semicircle of radius two has $\sigma_v^2=1$ and $\mu_4=2$.

\subsection{Hamiltonian commutator: semicircle self-convolution}
\label{app:convolution}

We use the manuscript convention $\rho_{\mathrm{sc},2}(u)=\frac{1}{2\pi}\sqrt{4-u^2}\,\mathbf 1_{|u|\le 2}$, where the subscript denotes the support radius. Define the difference density by
\begin{equation}
 \rho_{\Delta}(\omega)=\int_{\mathbb R}\rho_{\mathrm{sc},2}(u)\rho_{\mathrm{sc},2}(u-\omega)\,du,
 \qquad |\omega|\le4.
 \label{eq:delta_density_semicircle}
\end{equation}
We denote this density by $\rho_{\Delta,2}$, following the manuscript's notation. For independent semicircular eigenvalues $\varepsilon,\varepsilon'$, the difference $\omega=\varepsilon-\varepsilon'$ has this density. The empirical pair-difference measure $N^{-2}\sum_{ij}\delta_{\varepsilon_i-\varepsilon_j}$ converges to the same convolution when the empirical eigenvalue law converges to the semicircle; finite-$N$ level repulsion affects local structure, not this limiting density. The $N$ diagonal pairs have total mass $1/N$ at zero order.

The convolution integral can be evaluated explicitly in terms of complete
elliptic integrals. Writing $s=|\omega|$ and
$
m=\left(\frac{4-s}{4+s}\right)^2,
$
we obtain
\begin{equation} \nonumber
\rho_{\Delta,2}(\omega)
=
\frac{4+s}{24\pi^2}
\left[
(16+s^2)E(m)-8sK(m)
\right],
~0\le s\le 4,
\label{eq:semicircle_self_convolution}
\end{equation}
and $\rho_{\Delta,2}(\omega)=0$ for $|\omega|>4$. Here
$K(m)$ and $E(m)$
are the complete elliptic integrals of the first and second kind~\cite{GradshteynRyzhik}, respectively. At the origin, the expression is understood by continuity,
giving
$
\rho_{\Delta,2}(0)=\frac{8}{3\pi^2}
$.

\subsection{Complex diagonal jumps and Gaussian approximation}
For complex $v_m,v_m'$ define $q_m=v_m\overline{v_m'}-(|v_m|^2+|v_m'|^2)/2$, and the real two-vector $\mathbf t_m=(\operatorname{Re}q_m,\operatorname{Im}q_m)^T$. Then $(\operatorname{Re}d,\operatorname{Im}d)^T=\sum_m\gamma_m\mathbf t_m$. For identically distributed independent channels its mean and covariance are
\begin{equation}
 \boldsymbol\mu_D=\Gamma_D\,\mathbb E\mathbf t,
 \qquad \Sigma_D=\left(\sum_m\gamma_m^2\right)\operatorname{Cov}(\mathbf t).
\end{equation}
Under a multivariate Lindeberg condition, the centered, covariance-normalized sum approaches a bivariate normal distribution, with density $\rho_D(\mathbf u)=(2\pi\sqrt{\det\Sigma_D})^{-1}\exp[-(\mathbf u-\boldsymbol\mu_D)^T\Sigma_D^{-1}(\mathbf u-\boldsymbol\mu_D)/2]$, provided $\Sigma_D$ is nonsingular. Writing $\mathbf u=(x,y)^T$, $\mu_D=(\mu_x,\mu_y)^T$, and $\Sigma_D=\begin{psmallmatrix}\sigma_x^2&\sigma_{xy}\\\sigma_{xy}&\sigma_y^2\end{psmallmatrix}$, its conditional imaginary-part law at fixed $x$ has mean $m_y(x)=\mu_y+(\sigma_{xy}/\sigma_x^2)(x-\mu_x)$ and variance $s^2=\sigma_y^2-\sigma_{xy}^2/\sigma_x^2$. Thus the full density can equivalently be evaluated as
\begin{equation}
\begin{aligned}
\rho_{\mathcal L}(x+iy)
&=
\frac{e^{-(x-\mu_x)^2/(2\sigma_x^2)}}{\sqrt{2\pi\sigma_x^2}}
\\
&\quad\times
\int_{-4}^{4}
\frac{e^{-[y+\alpha\omega-m_y(x)]^2/(2s^2)}}{\sqrt{2\pi s^2}}
\rho_{\Delta,2}(\omega)\,d\omega .
\end{aligned}
\end{equation}

\section{Quenched-$A$ matrix Dyson equation}
\label{app:EC_MDE}\label{sec:appendix_A_Quenched}

Here we give the conditional matrix Dyson equation used in Sec.~\ref{sec:ensembleC_boundary}. The random-matrix model is
\begin{equation}
 X=Y+D,\qquad
 D=W_A\otimes\mathbbm1_B+\mathbbm1_A\otimes W_B,
 \label{eq:EC_app_model}
\end{equation}
where $Y$ is a complex Ginibre matrix of size $N_A N_B$ with entry variance $v_Y/(N_A N_B)$. The independent elliptic matrices $W_A$ and $W_B$ have the same distribution as $W=C/2+i\alpha H$ and obey
\begin{equation}
\begin{split}
 \mathbb E\,W_{ij}\overline{W_{kl}}
 &=\frac{\sigma^2}{N}\delta_{ik}\delta_{jl},\\
 \mathbb E\,W_{ij}W_{kl}
 &=\frac{c}{N}\delta_{il}\delta_{jk},
\end{split}
 \label{eq:EC_app_covariance}
\end{equation}
Here $\sigma^2=1/4+\alpha^2$ and $c=1/4-\alpha^2$. As in the main text, the two copies are taken independent for the calculation. The results below concern the iterated large-dimension limit, first $N_B\to\infty$ with $N_A$ and $W_A$ fixed, followed by $N_A\to\infty$.

We use the conditional expectation of the main text, acting blockwise
onto $\mathfrak B=\mathbb M_2(\mathbb C)\otimes\mathbb M_{N_A}(\mathbb C)$:
\begin{equation}
\begin{aligned}
\mathbb E_{\mathfrak B}[T]
&:=\mathbb E_{Y,W_B}\!\left[
(\operatorname{id}_A\otimes\operatorname{tr}_B)T
\right],\\
\operatorname{tr}_B&=\frac1{N_B}\operatorname{Tr}_B.
\end{aligned}
\label{eq:conditional_expectation_restored}
\end{equation}
Here $W_A$ is fixed, and an element of $\mathfrak B$ is identified
with its tensor product with $\mathbbm1_B$ when acting on the full space.
The Ginibre covariance used below is
\begin{equation}
\begin{aligned}
\mathbb E[Y_{IJ}\overline{Y}_{KL}]
&=\frac{v_Y}{N_A N_B}\delta_{IK}\delta_{JL},\\
\mathbb E[Y_{IJ}Y_{KL}]&=0.
\end{aligned}
\label{eq:Ycov}
\end{equation}

\subsection{Conditional resolvent and self-energy}

Conditionally on $W_A$, the matrix $\mathsf W_A$ belongs to
 the retained coefficient algebra and satisfies
 $\mathbb E_{\mathfrak B}[\mathsf W_A\otimes\mathbbm1_B]=\mathsf W_A$.
It therefore plays the role of the element $x\in\mathfrak B$ in
Speicher's operator-valued Pastur equation \cite{Speicher2011}. 
\begin{equation}
\begin{aligned}
\mathsf Q_A&=
\begin{pmatrix}
z\mathbbm1_A&iw\mathbbm1_A\\
iw\mathbbm1_A&\overline z\mathbbm1_A
\end{pmatrix}, \qquad
\mathsf W_A=
\begin{pmatrix}
W_A&0\\
0&W_A^\dagger
\end{pmatrix},
\\
\Kk_0&=\mathsf Q_A-\mathsf W_A.
\end{aligned}
\label{eq:EC_app_QA}
\end{equation}
and collect the remaining hermitized parts as 
\begin{equation}
\begin{aligned}
\sh&=\Yh+\xih,
 \qquad\Yh=
\begin{pmatrix}
Y&0\\
0&Y^\dagger
\end{pmatrix},
\\
\xih&=
\begin{pmatrix}
\mathbbm1_A\otimes W_B&0\\
0&\mathbbm1_A\otimes W_B^\dagger
\end{pmatrix}.
\end{aligned}
\label{eq:shat}
\end{equation}
The matrix $\Kk_0$ acts on the retained $2N_A$-dimensional space.
On the full $2N_AN_B$-dimensional space, we write its extension
explicitly as $\Kk_0\otimes\mathbbm1_B$.  Writing $\widehat X=\operatorname{diag}(X,X^\dagger)$, the finite-dimensional
resolvent and its conditional expectation are
\begin{equation}
\begin{aligned}
\mathsf R_X(z,w)&:=(\Kk_0\otimes\mathbbm1_B-\sh)^{-1},
\\
\GNA(z,w;W_A)&:=\mathbb E_{\fB}[\mathsf R_X]
\\
&=
\mathbb E_{\fB}\!\left[
\left(\mathsf Q_A\otimes\mathbbm1_B-\Xh\right)^{-1}
\right].
\end{aligned}
\label{eq:EC_app_conditional}
\end{equation}
Applying $\mathbb E_{\fB}$ to the exact resolvent identity of the form
$(\Kk_0 \otimes\mathbbm1_B
-\sh)\mathsf R_X=\mathbbm1_{2AB}$ gives
\begin{equation}
\Kk_0\GNA
=\mathbbm1_{2A}+\mathbb E_{\fB}[\sh \mathsf R_X].
\label{eq:exactunclosed}
\end{equation}
Equation~\eqref{eq:exactunclosed} is exact at finite $N_A,N_B$, but it is
not closed, since its last term still contains the full random resolvent
$\mathsf R_X$.

We denote the operator-valued large-$N_B$ limit, taken at fixed $N_A$ and
$W_A$, by
\begin{equation}
\begin{aligned}
\GA(z,w;W_A)&=\lim_{N_B\to\infty}\GNA(z,w;W_A),\\
\GA&=(G_{ab})_{a,b=1}^{2},
\end{aligned}
\end{equation}
and recall the normalized trace notation, which holds $q(m):=\frac1{N_A}\Tr_A m $ for  $ m\in \mathbb M_{N_A}(\mathbb C)$. The two independent random contributions in \eqref{eq:shat} generate two
covariance maps.  Evaluated at the unknown conditional Green function,
the Ginibre covariance map defined as 
\begin{equation}
\mathcal S_Y[\GA]
:=\mathbb E_{\fB}\!\left[
\Yh(\GA\otimes\mathbbm1_B)\Yh
\right].
\end{equation}
Its diagonal blocks vanish because
$\E[YMY]=\E[Y^\dagger M Y^\dagger]=0$.  In the $(1,2)$ block, using the Ginibre covariance \eqref{eq:Ycov}, one obtains
\begin{align}
\E_Y\!\left[Y(G_{12}\otimes\mathbbm1_B)Y^\dagger\right]
&=\frac{v_Y}{N_AN_B}
\Tr_{AB}(G_{12}\otimes\mathbbm1_B)\,\mathbbm1_{AB} \nonumber \\
&=v_Yq(G_{12})\mathbbm1_{AB}.
\end{align}
The last step uses the definition of $q$. After the normalized $B$-trace, this yields
\begin{equation}
\mathcal S_Y[\GA]
=v_Y
\begin{pmatrix}
0&q(G_{12})\mathbbm1_A\\
q(G_{21})\mathbbm1_A&0
\end{pmatrix}.
\label{eq:etaY}
\end{equation}
Thus the Ginibre channel traces out the retained $A$-structure and
returns a scalar multiple of $\mathbbm1_A$. The covariance map associated with the $B$-subspace is
\begin{equation}
\mathcal S_B[\GA]
=\mathbb E_{\fB}\!\left[
\xih(\GA\otimes\mathbbm1_B)\xih
\right].
\end{equation}
For example, in the $(1,1)$ block, on the element level we have
\begin{align}
&\left[
(\mathbbm1_A\otimes W_B)
(G_{11}\otimes\mathbbm1_B)
(\mathbbm1_A\otimes W_B)
\right]_{(ia)(jb)}\\
&=(G_{11})_{ij}(W_B^{\,2})_{ab},
\end{align}
and after averaging, this contribution reads
\begin{align}
\E(W_B^{\,2})_{ab}
=\sum_r
\E[(W_B)_{ar}(W_B)_{rb}]
=c\delta_{ab},
\end{align}
whereas in the $(1,2)$ block the situation is similar with 
\begin{equation}
\E(W_BW_B^\dagger)_{ab}
=\sigma^2\delta_{ab}.
\end{equation}
The $A$-matrix therefore passes through this contraction unchanged, and
\begin{equation}
\mathcal S_B[\GA]
=
\begin{pmatrix}
cG_{11}&\sigma^2G_{12}\\
\sigma^2G_{21}&cG_{22}
\end{pmatrix}.
\label{eq:EC_app_selfenergies}
\end{equation}
The asymmetry between \eqref{eq:etaY} and \eqref{eq:EC_app_selfenergies} is the origin
of the matrix character of the quenched equation: $\mathcal S_Y$ traces out the $A$ structure, and 
$\mathcal S_B$  preserves the full $A$-matrix structure.
Crossing Wick pairings of the $B$-leg are suppressed after the normalized $B$-trace. If $\kappa$ denotes either of the two relevant covariance coefficients,
$\kappa\in\{c,\sigma^2\}$, a representative fourth-order
crossing scales \cite{Shlyakhtenko1998,Speicher2011} as
\begin{equation}
\begin{aligned}
&\frac1{N_B}\sum_{a,b,r,d}
\frac{\kappa}{N_B}\delta_{ad}\delta_{br}
\frac{\kappa}{N_B}\delta_{ba}\delta_{rd}=\frac{\kappa^2}{N_B^3}\sum_a1
=\frac{\kappa^2}{N_B^2}.
\end{aligned}
\end{equation}
The Kronecker deltas force $a=b=r=d$ and leave only one free
$B$-index loop. In contrast, non-crossing pairings close the maximal
number of $B$-index loops and survive at order one. Mixed
$Y$--$W_B$ contractions vanish by centering and independence.

Consequently, in the iterated limit $N_B\to\infty$ at fixed $N_A$ and
$W_A$, only the non-crossing Wick pairings contribute to the
$\mathfrak B$-valued moments of the doubled random part.  The resulting
Gaussian closure is determined by the covariance map
$\mathcal S=\mathcal S_Y+\mathcal S_B$~\cite{Speicher2011}.
Equivalently, Gaussian integration by parts together with conditional
factorization closes the last term of Eq.~\eqref{eq:exactunclosed} as
$\mathbb E_{\fB}[\sh\mathsf R_X]=\mathcal S[\GA]\GA+o_{N_B}(1)$.
Taking $N_B\to\infty$ therefore gives,
\begin{equation}
\Kk_0\GA
=
\mathbbm1_{2A}+\mathcal S[\GA]\GA.
\end{equation}
Since $\Kk_0=\mathsf Q_A-\mathsf W_A$, this is equivalent to
\begin{equation}
\GA(z,w;W_A)
=
\left[
\mathsf Q_A-\mathsf W_A
-\mathcal S\bigl(\GA(z,w;W_A)\bigr)
\right]^{-1}.
\label{eq:EC_app_MDE}
\end{equation}

In Speicher's notation, the operator-valued semicircular element $s$
corresponds, under the standard self-adjoint block hermitization, to the
centered random part represented here by $\sh$.  The element $x$ is the
fixed quenched block $\mathsf W_A$, while the spectral argument is
$b=\mathsf Q_A$.  Speicher's covariance map
$\eta[T]=\mathbb E_{\fB}[sTs]$ corresponds to
$\mathcal S=\mathcal S_Y+\mathcal S_B$.  Equation~(22.7.2) of
Ref.~\cite{Speicher2011} then reads
\begin{equation}
G_{x+s}(b) =\left[
b-x-\eta\bigl(G_{x+s}(b)\bigr)
\right]^{-1}.
\end{equation}
After transforming back to the present quaternionic-block convention,
this is Eq.~\eqref{eq:EC_app_MDE}, with the identifications
\begin{equation}
\begin{aligned}
x&\longleftrightarrow\mathsf W_A,\\
b&\longleftrightarrow\mathsf Q_A,\\
G_{x+s}(b)&\longleftrightarrow\GA(z,w;W_A),\\
\eta &\longleftrightarrow\mathcal S_Y+\mathcal S_B.
\end{aligned}
\end{equation}
With $q_{ab}=q(G_{ab})=\frac1{N_A}\Tr_A G_{ab}$ the equation becomes explicitly 
\begin{widetext}
\begin{equation}
 \begin{pmatrix}
 G_{11} & G_{12} \\
 G_{21} & G_{22}
 \end{pmatrix} =
\begin{pmatrix}
z\mathbbm1_A-W_A-cG_{11}
&iw\mathbbm1_A-v_Yq_{12}\mathbbm1_A-\sigma^2G_{12}\\[2mm]
iw\mathbbm1_A-v_Yq_{21}\mathbbm1_A-\sigma^2G_{21}
&\overline z\mathbbm1_A-W_A^\dagger-cG_{22}
\end{pmatrix}^{-1}.
\label{eq:ensembleC_quenchedMDE}
\end{equation}
\end{widetext}
The inverse in \eqref{eq:ensembleC_quenchedMDE} is taken in the full
$2N_A$-dimensional block space; it is not a collection of four
independent inversions.
Only after solving \eqref{eq:ensembleC_quenchedMDE} at fixed $W_A$ do we take the
remaining normalized trace and the outer average:
\begin{equation}
\Gfin(z,w)
:=\E_{W_A}\!\left[
\frac1{N_A}\Tr_A\GA(z,w;W_A)
\right].
\label{eq:Gfinal}
\end{equation}
The trace in \eqref{eq:Gfinal} is taken blockwise; explicitly,
\begin{align}
[\Gfin(z,w)]_{ab}
&=\E_{W_A}\!\left[
\frac1{N_A}\sum_{i=1}^{N_A}
(G_{ab}(z,w;W_A))_{ii}
\right] \nonumber \\
&=\E_{W_A}[q_{ab}(W_A)].
\end{align}
Thus the order of operations is
\begin{equation}
\begin{aligned}
\text{fix/sample }W_A
&\ \longrightarrow\
\text{solve the matrix equation}
\\
&\ \longrightarrow\
\text{take }q
\ \longrightarrow\
\E_{W_A}.
\end{aligned}
\end{equation}
The second limit $N_A\to\infty$ is taken only after this procedure.
Feeding $\E_{W_A}q_{ab}$ back into \eqref{eq:ensembleC_quenchedMDE} would erase the
$A$-resolved structure that quenching was introduced to preserve.
\subsection{Derivation of the master boundary equation}
\label{sec:appendix_master_derivation}

The edge is the point at which the holomorphic solution first becomes unstable against an infinitesimal
off-diagonal perturbation. By setting $w=0$ we consider the solution with $G_{12}=G_{21}=0$. The diagonal blocks then decouple, with denoting the $(1,1)$ block by
\begin{equation}
M=M(z;W_A)
=\left[z\mathbbm{1}_A-W_A-cM\right]^{-1}.
\label{eq:EC_app_M_holo}
\end{equation}
The physical branch is selected by ~$M(z;W_A)\sim z^{-1}\mathbbm{1}_A
\qquad(|z|\to\infty)$.
Because $c\in \mathbb{R}$, the $(2,2)$ block on this branch is
$G_{22}=M^\dagger$.

Write the inverse matrix in \eqref{eq:ensembleC_quenchedMDE} as
$\Kk:=\Kk_0-\mathcal S[\GA]$, so that $\GA=\Kk^{-1}$.
For every invertible matrix the following identity holds,
\begin{equation}
\delta(\Kk^{-1})
=-\Kk^{-1}(\delta \Kk)\Kk^{-1}.
\label{eq:inversevariation}
\end{equation}
At the holomorphic point, $\Kk$ is block diagonal and
$(\Kk^{-1})_{11}=M,
$ and $
(\Kk^{-1})_{22}=M^\dagger.$
The $(1,2)$ entry of $\Kk$ is
\begin{equation}
\Kk_{12}
=-v_Yq_{12}\mathbbm{1}_A-\sigma^2G_{12},
\end{equation}
and therefore
$\delta \Kk_{12}
=-v_Y\delta q_{12}\mathbbm{1}_A-\sigma^2\delta G_{12},
$ with $
\delta q_{12}=q(\delta G_{12})$.
Taking the $(1,2)$ block of \eqref{eq:inversevariation}, we have
\begin{align}
\delta G_{12}
&=-M(\delta \Kk_{12})M^\dagger =M\left[v_Y\delta q_{12}\mathbbm{1}_A
+\sigma^2\delta G_{12}\right]M^\dagger.
\label{eq:EC_app_linearization}
\end{align}
This explains both minus signs: one comes from differentiating the inverse,
the other from the fact that the self-energy is subtracted in
$\Kk=\mathsf Q_A-\mathsf W_A-\mathcal S(\GA)$.
To resum the resulting series, recall the definition of the linear map on $\mathbb M_{N_A}(\mathbb{C})$,
\begin{equation}
\Lambda_z[\Delta]:=\sigma^2M\Delta M^\dagger.
\label{eq:Lmap}
\end{equation}
Then \eqref{eq:EC_app_linearization} becomes
\begin{equation}
(1-\Lambda_z)[\delta G_{12}]
=v_Y\delta q_{12}MM^\dagger.
\end{equation}
Whenever $1-\Lambda_z$ is invertible, we can write $\delta G_{12}$ in the form
\begin{equation}
\delta G_{12}
=v_Y\delta q_{12}
(1-\Lambda_z)^{-1}[MM^\dagger].
\label{eq:EC_app_offdiag}
\end{equation}
Now we take the normalized $A$-trace.  Since
$\delta q_{12}=q(\delta G_{12})$, where we recall $q= \frac{1}{N_{A}}\Tr_{A}$, we obtain an equation for the variation of $q_{12}$:
\begin{equation}
\delta q_{12}
=v_Y\delta q_{12}
q\!\left((1-\Lambda_z)^{-1}[MM^\dagger]\right).
\label{eq:dqequation}
\end{equation}
Outside the spectrum the only stable solution is $\delta q_{12}=0$. At the boundary a non-zero infinitesimal solution bifurcates.  Cancelling
$\delta q_{12}\neq0$ in Eq. ~\eqref{eq:dqequation} gives
\begin{equation}
q\!\left((1-\Lambda_z)^{-1}[MM^\dagger]\right)=\frac1{v_Y}.
\label{eq:operatorboundary}
\end{equation}
Let $\operatorname{spr}(\Lambda_z)$ denote the spectral radius of
the linear operator $\Lambda_z$.
If $\operatorname{spr}(\Lambda_z)<1$, the inverse has the Neumann
expansion 
$(1-\Lambda_z)^{-1}=\sum_{k=0}^\infty\Lambda_z^k.$
Repeated application of \eqref{eq:Lmap} gives
\begin{equation}
\Lambda_z^k[MM^\dagger]
=(\sigma^2)^kM^{k+1}(M^\dagger)^{k+1}.
\end{equation}
Therefore we can rewrite Eq.~\eqref{eq:operatorboundary} as
\begin{align}
q\!\left((1-\Lambda_z)^{-1}[MM^\dagger]\right)
&=\sum_{k=0}^\infty(\sigma^2)^k
q\!\left(M^{k+1}(M^\dagger)^{k+1}\right) \nonumber \\
&=\sum_{n=1}^\infty(\sigma^2)^{n-1}
q\!\left(M^n(M^\dagger)^n\right).
\end{align}
Here $q(T)=N_A^{-1}\operatorname{Tr}_A T$ is the normalized trace at
fixed $N_A$ and fixed $W_A$, whereas
$\varphi(T)=\lim_{N_A\to\infty}\mathbb E_{W_A}[q(T)]$ denotes its
averaged large-$N_A$ limit.

Let $M$ be the branch defined in Eq.~\eqref{eq:ensembleC_holomorphic_def}.  
We use the function $\Phi(z)$ defined in Eq.~\eqref{eq:ensembleC_phi_def}, 
\begin{equation}
\begin{aligned}
v_Y\Phi(z)&=1,\\
\Phi(z)
&=\sum_{n=1}^{\infty}(\sigma^2)^{n-1}
\varphi\!\left[
M(z)^n\bigl(M(z)^\dagger\bigr)^n
\right].
\end{aligned}
\label{eq:ensembleC_master_boundary}
\end{equation}
On the exterior holomorphic branch, the boundary of the quenched-$A$
solution is determined by $\Phi(z)=1/v_Y$.
Note that the operator form \eqref{eq:operatorboundary} is the primary
statement; the series \eqref{eq:ensembleC_master_boundary} is its Neumann expansion and is
valid where
\begin{equation}
\operatorname{spr}(\Lambda_z)=\sigma^2\operatorname{spr}(M)^2<1.
\label{eq:EC_app_Stein_domain}
\end{equation}  
The conditional MDE is first solved at fixed $W_A$.  The resulting
normalized-trace quantity is then averaged over $W_A$, as in
Eqs.~\eqref{eq:Gfinal} and~\eqref{eq:ensembleC_master_boundary}, and the
boundary is obtained from $v_Y\Phi(z)=1$.  At large $N_A$, the
finite-sample quantity entering this average is self-averaging; averaging
over several realizations reduces the finite-$N_A$ ripple but is not fed
back into the conditional MDE. 
For numerical evaluation, it is preferable to resum \eqref{eq:ensembleC_master_boundary} instead of truncating it term by term.  Define the
matrix
\begin{equation}
Z=\sum_{k=0}^{\infty}
(\sigma^2)^kM^k(M^\dagger)^k.
\label{eq:Stein_series}
\end{equation}
Multiplying \eqref{eq:Stein_series} by $\sigma^2M$ from the left and
by $M^\dagger$ from the right shifts the summation index and gives
\begin{equation}
\sigma^2MZM^\dagger
=\sum_{k=1}^{\infty}(\sigma^2)^kM^k(M^\dagger)^k
=Z-\mathbbm{1}_A.
\end{equation}
Hence $Z$ is equivalently obtained from the discrete Lyapunov, or
Stein, equation
\begin{equation}
Z=\mathbbm{1}_A+\sigma^2MZM^\dagger
=\mathbbm{1}_A+\Lambda_z[Z].
\label{eq:ensembleC_Stein_numerical}
\end{equation}
Taking the normalized trace of Eq.~\eqref{eq:Stein_series}, using
$q(\mathbbm{1}_A)=1$, and averaging over $W_A$, we obtain
\begin{equation}
\begin{aligned}
\mathbb E_{W_A}\!\left[q(Z)\right]
&=1+\sum_{k=1}^{\infty}(\sigma^2)^k
\mathbb E_{W_A}\!\left[q\!\left(M^k(M^\dagger)^k\right)\right].
\end{aligned}
\end{equation}
Therefore, in the limit of large-$N_A$ one obtains,
\begin{equation}
\Phi(z)
=\frac{\varphi(Z)-1}{\sigma^2}.
\label{eq:Phi_Stein}
\end{equation}
This avoids the critical slowing of iterating the full non-holomorphic
$2N_A\times2N_A$ system.  The condition
$\operatorname{spr}(\Lambda_z)<1$ is required for the convergent
Neumann-series interpretation.  Outside this region an algebraic Stein
solution may still exist, but it need not be the positive response
associated with the exterior holomorphic branch.
The master equation~\eqref{eq:ensembleC_master_boundary} was obtained above
directly, as the loss of stability of the holomorphic branch of the
quenched-$A$ equation~\eqref{eq:ensembleC_quenchedMDE}.  It is worth recording that it can be recognised, a posteriori, as an instance of a
general criterion.  For a model of the form $X=Y+D$, in which $Y$ is
circular of variance $v_Y$ and free from $D$, the boundary of the support of
the Brown measure of $X$, restricted to the regular exterior of
$\operatorname{spec}D$, satisfies
\begin{equation}
v_Y\,\varphi\!\left[(D-z)^{-1}\bigl((D-z)^{-1}\bigr)^{\dagger}\right]=1 ;
\label{eq:FZ_general}
\end{equation}
for Hermitian $D$ this is the Feinberg--Zee condition underlying, after a
$90^\circ$ rotation of the spectral plane, the treatment of Ensemble~B in
Sec.~\ref{ensembleB}, and in the present setting it is covered by the recent analyses of
Refs.~\cite{zhong2025brownmeasuresumelliptic, erdos2024densitybrownmeasurefree,ho2021brownmeasuresfreecircular}.
We stress that the applicability of Eq~\eqref{eq:FZ_general} is not affected by the tensor structure: the Ginibre part $Y$ is bi-unitarily invariant and independent of $W_A,W_B$, hence asymptotically free from $D=W_A\otimes\mathbbm 1_B+\mathbbm 1_A\otimes W_B$ irrespective of how $D$ itself is built.  It is only the two legs of $D$ that fail to be free from each other.

The criterion is, however, only as useful as one's access to the
two-resolvent functional on its left-hand side, and for a tensor sum of
elliptic elements this quantity is not determined by the assumption of freeness between the tensor factors.  Since the two legs commute, their $*$-distribution is a classical and not a free
combination, and replacing them by freely independent variables gives incorrect results: the
scalar closure of  
the covariance-matched elliptic Gaussian approximation of Appendix ~\ref{app:EC_scalar} gives the
ellipse~\eqref{eq:EC_app_scalar_ellipse}, the
same free-addition approximation in terms of the quaternionic $R$-transform
gives the same ellipse, and applying the
Haagerup--Larsen formula~\cite{haageruplarsen2000} outside its $R$-diagonal setting
gives $\sqrt2$ at
$\alpha=1/2$, against the correct value~\eqref{eq:ensembleC_silver_radius}.

What the quenched-$A$ construction supplies is precisely a computation
of this functional.  Keeping one tensor factor inside the self-consistency equation and integrating out the $B$ factor gives
\begin{equation}
\Phi(z)=\varphi\!\left[(D-z)^{-1}\bigl((D-z)^{-1}\bigr)^{\dagger}\right],
\label{eq:Phi_is_FZ}
\end{equation}
Thus Eqs.~\eqref{eq:ensembleC_master_boundary}
and~\eqref{eq:FZ_general} give the same boundary condition.  For $v_Y=1$,
$\Phi(z)=1$ corresponds to Eq.~(3.3) of
Ref.~\cite{zhong2025brownmeasuresumelliptic}; in the present approach, the
required function is evaluated by the Neumann series, equivalently by the
Stein equation~\eqref{eq:ensembleC_Stein_numerical}. Appendix~\ref{sec:appendix_C_lemon} provides a check of the
identification~\eqref{eq:Phi_is_FZ}: at $\alpha=0$ the element $D$ is
Hermitian, the two-resolvent functional in \eqref{eq:FZ_general} reduces to
$-\operatorname{Im}\GDl{1}(z)/\operatorname{Im}z$, and the criterion returns the
lemon contour~\eqref{eq:lemon_contour}.

Read in this way, Ensembles~B and~C are two instances of one equation,
differing only in the element $D$ to which it is applied: for Ensemble~B,
$D$ is the anti-Hermitian Hamiltonian contribution in Eq.~\eqref{eq:ilemon}
and becomes Hermitian after a $90^\circ$ rotation; for Ensemble~C, matrix $D$ is the tensor sum of
two elliptic elements, Hermitian with density $\rD{1}$ at $\alpha=0$ and
non-normal otherwise.  Only in the latter case is the input to the criterion
not directly available, and it is this gap that the quenched-$A$ equation
fills.

\subsection{Spectral density}
\label{app:quenchedA_density}

The planar density, including the isotropic radial density in Fig.~\ref{fig:ensembleC_radial}, is instead recovered from the non-holomorphic $w>0$ solution:
\begin{equation}
 \rho_X(z)=\frac1\pi\partial_{\overline z}
 \lim_{w\to0^+}\lim_{N_A\to\infty}
 \mathbb E_{W_A}\,q_{11}(z,w;W_A).
 \label{eq:ensembleC_density_general}
\end{equation}
Thus the boundary calculation and the density calculation employ different branches of the same hermitized equation.

\subsection{Finite-dimensional implementation of the quenched-$A$ equation}
\label{app:quenchedA_convergence}

The quenched-$A$ matrix Dyson equation involves a normalized trace over the $A$ space followed by an average over $W_A$,
\[
g(r)=
\mathbb{E}_{W_A}\!\left[
\frac{1}{N_A}\operatorname{Tr}G_{11}(r;W_A)
\right].
\]
In the analytical construction, this quantity is understood in the limit $N_A\to\infty$. In numerical calculations this limit is approximated by a finite Ginibre matrix $W_A$ of dimension $N_A$. We emphasize that $N_A$ is therefore an auxiliary numerical parameter and should not be identified with the physical Hilbert-space dimension $N$ used in the direct sampling of Lindbladian spectra.

The convergence with $N_A$ is strongly nonuniform across the spectrum. In the bulk, the normalized trace is strongly self-averaging and moderate values of $N_A$ already give a good approximation to the limiting solution. Close to the spectral boundary, however, the holomorphic solution approaches its stability threshold. Schematically, the response to a finite-$N_A$ perturbation takes the form
\[
\delta G\simeq
(1-\mathcal{K}_r)^{-1}\delta G_{N_A},
\]
where $\mathcal{K}_r$ is the linearized stability operator. At the spectral edge its leading eigenvalue approaches unity, so that the norm of $(1-\mathcal{K}_r)^{-1}$ becomes large. Consequently, small finite-$N_A$ fluctuations of the auxiliary Ginibre realization are strongly amplified near the edge. This explains why substantially larger values of $N_A$ are required to resolve accurately the decreasing tail of the radial density than to reproduce its bulk.

A second amplification arises when the Green function is converted into the radial density. At the balanced point,
\[
F(r)=r\,\operatorname{Re}g(r),
\qquad
\rho(r)
=
\frac{1}{2\pi r}
\frac{d}{dr}\left[r\,\operatorname{Re}g(r)\right].
\]
Hence small residual fluctuations of $g(r)$ are differentiated and become more visible in $\rho(r)$, particularly close to the boundary. In the numerical implementation we therefore differentiate the quenched-averaged Green function only after averaging and use a local polynomial derivative; close to the exact edge the derivative is evaluated from interior points only.

The regularization parameter $w>0$ provides an additional source of edge rounding. At finite $w$ the non-analytic spectral boundary is smoothed and the regularized solution develops a small tail beyond the exact support. The physical density is recovered by taking the large-$N_A$ limit together with the subsequent small-$w$ limit,
\[
N_A\rightarrow\infty,
~~
w\rightarrow0.
\]
Reducing $w$ makes the calculation increasingly sensitive to finite-$N_A$ fluctuations because the solution approaches the singular stability threshold more closely. Thus, for a fixed small regulator, the auxiliary dimension required for accurate convergence near the edge can be substantially larger than the physical dimensions used in direct Monte Carlo sampling.

For this reason, the theoretical curve shown in Fig.~\ref{fig:ensembleC_radial} should be interpreted as an approximation to the asymptotic $N\to\infty$ quenched-$A$ result, obtained with a sufficiently large auxiliary dimension $N_A$. The values $N=100$ and $N=200$ quoted in the figure refer only to the direct finite-$N$ sampling of random Lindbladians and need not coincide with $N_A$. The raw finite-$w$ solution may retain a small regulator-induced tail outside the exact radius $\bar r=\sqrt{1+\sqrt{2}}$; in the support-enforced curve this tail is removed and the density is set to zero outside the analytically determined support.

\section{Elliptic two-resolvent kernel and the contour integral}
\label{app:EC_kernel}\label{sec:appendix_B_boundary}

We next derive the integral form of $\Phi(z)$ and the representation used for numerical quadrature. Let $g(\lambda)$ be the exterior scalar Cauchy transform of $W$. The Gaussian elliptic covariance gives
\begin{equation}
 \begin{split}
 g(\lambda)^{-1}&=\lambda-cg(\lambda),\\
 g(\lambda)&=\frac{2}{\lambda+\sqrt{\lambda^2-4c}}
 \sim\lambda^{-1}.
 \end{split}
 \label{eq:EC_app_g}
\end{equation}
The square root is chosen analytically so that $g(\lambda)\sim1/\lambda$. Likewise Eq.~\eqref{eq:EC_app_M_holo} yields $M=f_z(W_A)$, with
\begin{equation}
 \begin{split}
 f_z(\lambda)^{-1}&=z-\lambda-cf_z(\lambda),\\
 f_z(\lambda)&=\frac{2}{z-\lambda+
 \sqrt{(z-\lambda)^2-4c}}.
 \end{split}
 \label{eq:EC_app_f}
\end{equation}
The same exterior-branch convention applies. When $c=0$, $g(\lambda)=1/\lambda$ and $f_z(\lambda)=1/(z-\lambda)$.

\subsection{Reduction to a positive double series}

The necessary input is the two-resolvent kernel, not only the eigenvalue density of the elliptic matrix. For exterior arguments $\lambda$ and $\nu$ the planar Gaussian contractions give
\begin{equation}
 \begin{split}
 \mathcal K(\lambda,\nu)
 &=\varphi\!\left[(\lambda-W)^{-1}
 (\nu-W^\dagger)^{-1}\right]\\
 &=\frac{g(\lambda)g(\nu)}
 {1-\sigma^2g(\lambda)g(\nu)}.
 \end{split}
 \label{eq:ensembleC_elliptic_kernel}
\end{equation}
The parameter $c$ determines the one-resolvent functions, whereas $\sigma^2$ controls contractions between opposite matrix orientations. For example, when $c=0$ the kernel reduces to $(\lambda\nu-\sigma^2)^{-1}$.

Let $\Gamma$ enclose the spectral ellipse of $W_A$ in a domain where $f_z$ is analytic. Holomorphic functional calculus, followed by the kernel expansion, gives
\begin{equation}
 \begin{split}
 a_{nk}(z)&=\oint_{\Gamma}\frac{\mathrm d\lambda}{2\pi i}
 f_z(\lambda)^n g(\lambda)^{k+1},\\
 \varphi[M^n(M^\dagger)^n]
 &=\sum_{k=0}^{\infty}(\sigma^2)^k |a_{nk}(z)|^2.
 \end{split}
 \label{eq:EC_app_ank}
\end{equation}
Conjugate Cauchy integrals combine into the absolute squares. The geometric expansion is justified when $|\sigma^2g(\lambda)g(\nu)|<1$ uniformly on the two enclosing contours. Inserting Eq.~\eqref{eq:EC_app_ank} into the stability series~\eqref{eq:ensembleC_master_boundary} yields
\begin{equation}
 \Phi(z)=\sum_{n=1}^{\infty}(\sigma^2)^{n-1}
 \sum_{k=0}^{\infty}(\sigma^2)^k |a_{nk}(z)|^2.
 \label{eq:EC_app_positive_series}
\end{equation}
This is the positive contour representation used in the main text. The order of the two series and the contour integrals must not be changed without convergence control.

There is also a direct interpretation in terms of the original source operator $D$. For $z$ in the regular exterior of its spectrum, define
\begin{equation}
 \begin{split}
 \mathcal F_D(z)&=\lim_{N_A,N_B\to\infty}
 \frac1{N_A N_B}\mathbb E\operatorname{Tr}\,[R_DR_D^\dagger],\\
 R_D&=(z\mathbbm1-D)^{-1}.
 \end{split}
 \label{eq:EC_app_source_twores}
\end{equation}
In the same iterated limiting procedure as in Appendix~\ref{app:EC_MDE}, the $B$-factor Wick contractions give
\begin{equation}
 \mathcal F_D(z)=\sum_{n\ge1}(\sigma^2)^{n-1}
 \varphi[M^n(M^\dagger)^n]=\Phi(z)
 \label{eq:EC_app_source_phi}
\end{equation}
where both representations are admissible. This recovers the source-resolvent identity~\eqref{eq:Phi_is_FZ}. Thus the condition
$v_Y\Phi(z)=1$ is the regular-exterior version of the general
support criterion~\eqref{eq:FZ_general} for adding an independent circular operator
to $D$~\cite{zhong2025brownmeasuresumelliptic,erdos2024densitybrownmeasurefree}.
Importantly, Eq.~\eqref{eq:EC_app_source_phi} is not an analytic
continuation through the source spectrum. It makes no statement by
itself about possible contact between the boundaries of $D$ and $X$.

Where both geometric expansions converge uniformly, one may formally combine them into the double-contour integral
\begin{widetext}
\begin{equation}
 \Phi(z)=\oint_{\Gamma}\frac{\mathrm d\lambda}{2\pi i}
 \oint_{\overline\Gamma}\frac{\mathrm d\nu}{2\pi i}
 \frac{g(\lambda)g(\nu)}{1-\sigma^2g(\lambda)g(\nu)}
 \frac{f_z(\lambda)f_{\overline z}(\nu)}
 {1-\sigma^2f_z(\lambda)f_{\overline z}(\nu)}.
 \label{eq:ensembleC_resummed_integral}
\end{equation}
\end{widetext}
Here $\overline\Gamma$ denotes a suitably oriented contour around the conjugate ellipse. This rational representation can acquire poles on a deformed contour and is not used for the large-$\alpha$ numerics. No analytic continuation through such poles is assumed.

\subsection{Elliptic coordinate and Fourier coefficients}

The defining relation of $g$ can be inverted as $\lambda=g^{-1}+cg$. We use this inverse map to choose the integration contour. For $\sigma=\sqrt{\sigma^2}$ and $0<q<1$ set
\begin{equation}
 \begin{split}
 g(\lambda_q(\theta))&=\frac q\sigma e^{-i\theta},\\
 \lambda_q(\theta)&=\frac\sigma q e^{i\theta}
 +\frac{cq}{\sigma}e^{-i\theta},\qquad
 0\le\theta<2\pi.
 \end{split}
 \label{eq:EC_app_ellipse_param}
\end{equation}
At $q=1$ this curve is the boundary of the limiting spectrum of $W$, with real and imaginary semiaxes $(\sigma^2+c)/\sigma$ and $(\sigma^2-c)/\sigma$. For $q<1$ the ellipse surrounds that support. It is nevertheless admissible for the integral only when the selected branch of $f_z$ is analytic on the enclosed region.

Define the Fourier coefficients
\begin{equation}
 F_{nk}(z;q)=\frac1{2\pi}\int_0^{2\pi}
 f_z(\lambda_q(\theta))^n e^{-ik\theta}\,\mathrm d\theta.
 \label{eq:EC_app_Fnk}
\end{equation}
Since $\lambda_q'(\theta)=i[(\sigma/q)e^{i\theta}-(cq/\sigma)e^{-i\theta}]$, substitution into the contour integral for $a_{nk}$ gives exactly
\begin{equation}
 a_{nk}(z)=\left(\frac q\sigma\right)^k
 \left[F_{nk}(z;q)-\frac{cq^2}{\sigma^2}F_{n,k+2}(z;q)\right].
 \label{eq:EC_app_fourier_ank}
\end{equation}
Thus the stability function can be evaluated entirely from one-dimensional Fourier coefficients:
\begin{equation}
 \begin{split}
 \Phi(z)&=\sum_{n\ge1}(\sigma^2)^{n-1}
 \sum_{k\ge0}q^{2k}\left|B_{nk}(z;q)\right|^2,\\
 B_{nk}(z;q)&=F_{nk}(z;q)-\frac{cq^2}{\sigma^2}
 F_{n,k+2}(z;q).
 \end{split}
 \label{eq:ensembleC_Fourier}
\end{equation}
The positive form avoids cancellations between large contributions to the resummed double integral. It is independent of $q$ only when the deformation stays inside an admissible analytic domain and both sums converge.

\section{The purely dissipative limit, $\alpha=0$}
\label{app:EC_lemon}\label{sec:appendix_C_lemon}
At $\alpha=0$, the support reduces to the lemon, and our aim is simply to verify that the method reproduces the known result. All terms take the form
\begin{equation}
W_A=\frac12C_A,
\qquad
c=\sigma^2=\frac14.
\end{equation}
Diagonalize $W_A=U\operatorname{diag}(\varepsilon_i)U^\dagger$, with $U$ unitary.  Equation
\eqref{eq:ensembleC_holomorphic_def} closes on diagonal matrices in the same basis.  Its
entries $f_i=f(z,\varepsilon_i)$ satisfy
\begin{equation}
f=\frac1{z-\varepsilon-\frac14f}.
\label{eq:EC_app_lemon_f}
\end{equation}
The physical solution is the Stieltjes transform of a semicircle of radius
one,
\begin{equation}
f(z,\varepsilon)=\Gsc{1}(z-\varepsilon),
\qquad
\Gsc{1}(u)=2\left(u-\sqrt{u^2-1}\right),
\end{equation}
with $\Gsc{1}(u)=\int \frac{\rsc{1}(\lambda)}{u-\lambda} \dd\lambda$  and  the branch $\Gsc{1}(u)\sim1/u$.
Write $z=x+iy$ with $y\ne0$.  Taking the imaginary part of Eq.~\eqref{eq:EC_app_lemon_f} and rearranging gives the useful identity,
\begin{equation}
\frac{|f|^2}{1-\frac14|f|^2}
=-\frac{\operatorname{Im}f}{y}.
\label{eq:EC_app_lemon_identity}
\end{equation}
For this diagonal $M$, the boundary equation~\eqref{eq:ensembleC_master_boundary} for $\Phi(z)$ becomes a geometric series, which is summing together with the help of Eq.~\eqref{eq:EC_app_lemon_identity} to:
\begin{equation}
\begin{aligned}
\Phi(z)
&=\int\rsc{1}(\varepsilon)
\sum_{n=1}^\infty\left(\frac14\right)^{n-1}|f(z,\varepsilon)|^{2n}\,\dd\varepsilon
\\
&=\int\rsc{1}(\varepsilon)
\frac{|f(z,\varepsilon)|^2}{1-\frac14|f(z,\varepsilon)|^2}\,\dd\varepsilon\\
&=-\frac1y\int\rsc{1}(\varepsilon)
\operatorname{Im}f(z,\varepsilon)\,\dd\varepsilon\\
&= - \frac{\operatorname{Im}\GDl{1}(z)}{y}.
\end{aligned}
\label{eq:EC_app_lemon_series}
\end{equation}
The final integral is the Stieltjes transform of the classical convolution
\begin{equation}
\begin{aligned}
\rD{1}:=\rsc{1}*\rsc{1}, \quad
\GDl{1}(z):=\int_{-2}^2
\frac{\rD{1}(\ell)}{z-\ell}\,\dd\ell.
\end{aligned}
\end{equation}
Indeed, one semicircle enters through the quenched eigenvalue $\varepsilon$, while
the other has already been resummed in $f=\Gsc{1}(z-\varepsilon)$. For $v_Y=1$, the boundary condition $\Phi(z)=1$ becomes $\operatorname{Im}[z+\GDl{1}(z)]=0$, exactly as in Eq.~\eqref{eq:lemon_contour}.

The limiting form on the real axis follows by continuity from the regular exterior.

\section{The balanced point $\alpha=1/2$ and the silver radius}
\label{app:EC_silver}\label{sec:appendix_D_silver}
Because $c=0$, the holomorphic block in Eq~\eqref{eq:ensembleC_holomorphic_def} reads
\begin{equation}
M=(z\mathbbm{1}_A-W_A)^{-1}.
\end{equation}
Throughout this appendix $\varphi$ denotes the limiting normalized
trace, and $W_A$ is circular with variance
$\varphi(W_AW_A^\dagger)=\sigma^2$.  Writing $\zeta=|z|^2$ and
$\Phi(\zeta):=\Phi(z)$, define the terms of the series inside the boundary
equation~\eqref{eq:ensembleC_master_boundary} by
\begin{equation}
\begin{aligned}
S_n(\zeta)
&:=\varphi\!\left[
(z\mathbbm{1}_A-W_A)^{-n}
(\overline z\mathbbm{1}_A-W_A^\dagger)^{-n}
\right]\\
&=\varphi\!\left[M^n(M^\dagger)^n\right]
=\lVert M^n\rVert_2^2,\\
S_0(\zeta)&  :=1.
\end{aligned}
\label{eq:EC_app_isotropic_Sn}
\end{equation}
Inserting this form into the boundary equation gives
\begin{equation}
\begin{aligned}
\Phi(\zeta)
&=\sum_{n=1}^\infty (\sigma^2)^{n-1} S_n(\zeta)
=\frac1{v_Y}.
\end{aligned}
\label{eq:half}
\end{equation}
Rotational invariance is visible analytically: every $S_n(\zeta)$ depends only on $\zeta$, so every level set of $\Phi$ is a circle.
For a circular element $W_A$ of variance $\sigma^2$, we have
\begin{equation}
\varphi\!\left[W_A^m(W_A^\dagger)^n\right]=\delta_{mn} \sigma^{2n}.
\label{eq:circularmoments}
\end{equation}
For large $|z|$ the expansion of the two resolvents holds,
\begin{equation}
(z-W_A)^{-1}=\sum_{m=0}^\infty\frac{W_A^m}{z^{m+1}},
\qquad
(\overline z-W_A^\dagger)^{-1}
=\sum_{n=0}^\infty\frac{(W_A^\dagger)^n}{\overline z^{\,n+1}}.
\end{equation}
Using \eqref{eq:circularmoments} we compute the first term of the sum directly, and as we will see this is enough to obtain all the others,
\begin{align}
S_1(\zeta)
&=\sum_{m,n\ge0}
\frac{\varphi[W_A^m(W_A^\dagger)^n]}
{z^{m+1}\overline z^{\,n+1}}=\sum_{n=0}^\infty\frac{\sigma^{2n}}{\zeta^{n+1}}
=\frac1{\zeta-\sigma^2},
\end{align}
where the last step is a geometric series.
This is also the $c=0$ specialization of Eq.~\eqref{eq:ensembleC_elliptic_kernel}, with $\lambda=z$ and $\nu=\overline z$.
The two factors are separately holomorphic in $z$ and $\overline z$, so we first write their $n-1$ derivatives,
\begin{equation}
\partial_z^{n-1}(z-W_A)^{-1}
=(-1)^{n-1}(n-1)!(z-W_A)^{-n},
\end{equation}
\begin{equation}
\partial_{\overline z}^{n-1}(\overline z-W_A^\dagger)^{-1}
=(-1)^{n-1}(n-1)!(\overline z-W_A^\dagger)^{-n}.
\end{equation}
Repeated differentiation then allows one to compute the terms $S_n(\zeta)$ exactly, and in an elegant way, in terms of $\zeta$,
\begin{equation}
\begin{aligned}
S_n(\zeta)
&=\frac1{[(n-1)!]^2}
\partial_z^{n-1}\partial_{\overline z}^{n-1}
\frac1{z\overline z-\sigma^2}\\
&=\frac{1}{(\zeta-\sigma^2)^{2n-1}} \cdot
\sum_{k=0}^{n-1}
\binom{n-1}{k}^{\!2}\sigma^{2k}\zeta^{n-1-k}.
\end{aligned}
\label{eq:EC_app_isotropic_moments}
\end{equation}
For example, the first three terms are
$S_1(\zeta)=1/(\zeta-\sigma^2)$,
$S_2(\zeta)=(\zeta+\sigma^2)/(\zeta-\sigma^2)^3$, and
$S_3(\zeta)=(\zeta^2+4\sigma^2\zeta+\sigma^4)/(\zeta-\sigma^2)^5$.
Put $m=n-1$ in \eqref{eq:half}.  Substituting
\eqref{eq:EC_app_isotropic_moments}, we arrive at
\begin{equation}
\begin{aligned}
\Phi(\zeta)
=\sum_{m=0}^\infty
\frac{\sigma^{2m}}
{(\zeta-\sigma^2)^{2m+1}} \cdot
\sum_{k=0}^{m}\binom{m}{k}^{\!2}
\sigma^{2k}\zeta^{m-k}.
\end{aligned}
\end{equation}
The polynomial in the numerator is expressed through the Legendre
polynomial $P_m$:
\begin{equation}
\begin{aligned}
&\sum_{k=0}^{m}\binom{m}{k}^{\!2}\sigma^{2k}\zeta^{m-k}=(\zeta-\sigma^2)^mP_m\!\left(\frac{\zeta+\sigma^2}{\zeta-\sigma^2}\right).
\end{aligned}
\end{equation}
Using the generating function of the Legendre polynomials, $\sum_{m=0}^\infty P_m(x)s^m
=\frac1{\sqrt{1-2xs+s^2}}$, we use it inside the boundary equation,
\begin{equation}
\begin{aligned}
\Phi(\zeta)
&=\frac1{\zeta-\sigma^2}
\sum_{m=0}^\infty
P_m\!\left(\frac{\zeta+\sigma^2}{\zeta-\sigma^2}\right)
\left(\frac{\sigma^2}{\zeta-\sigma^2}\right)^m\\
&=\frac1{\sqrt{\zeta\left(\zeta-4\sigma^2\right)}},
\qquad \zeta>4\sigma^2.
\end{aligned}
\label{eq:EC_app_isotropic_phi}
\end{equation}
If the variances of $W_A$ and $W_B$ are kept formally distinct and
denoted by $v$ and $t$, respectively, the same computation gives
$\Phi(\zeta)=[(\zeta+v-t)^2-4\zeta v]^{-1/2}$; the expression above is
recovered by setting $v=t=\sigma^2$.
Thus, at the balanced point $\alpha=1/2$, rotational invariance makes
the condition $\Phi(\zeta)=1/v_Y$ a circle.
Thus $v_Y\Phi(\zeta)=1$ yields
\begin{equation}
 \zeta=2\sigma^2+\sqrt{4\sigma^4+v_Y^2}.
 \label{eq:EC_app_isotropic_radius_general}
\end{equation}
For the present normalization $v_Y=1$ and $\sigma^2=1/2$, the support of the model is a disk with radius
\begin{equation}
\bar r=\sqrt{1+\sqrt2}\simeq1.553774.
 \label{eq:EC_app_silver_radius}
\end{equation}


\section{Why a scalar closure of the Dyson equation does not apply}
\label{app:EC_scalar}\label{sec:appendix_E_scalarMDE}

A tempting approximation is to treat $X$ as a Gaussian non-Hermitian random matrix characterized only by its two scalar covariance moments. For the independent-copy tensor model, these are
\begin{equation}
 c_o=\varphi(XX^\dagger)=v_Y+2\sigma^2,
 \qquad c_d=\varphi(X^2)=2c.
 \label{eq:EC_app_scalar_moments}
\end{equation}
Replacing the full tensor structure by a second-moment-matched elliptic Gaussian approximation produces the ellipse
\begin{equation}
 \frac{x^2}{a^2}+\frac{y^2}{b^2}=1,
 \qquad a=\frac{c_o+c_d}{\sqrt{c_o}},\quad
 b=\frac{c_o-c_d}{\sqrt{c_o}}.
 \label{eq:EC_app_scalar_ellipse}
\end{equation}
At $\alpha=0$ it predicts $(a,b)=(4/\sqrt6,2/\sqrt6)$ rather than the lemon contour. At $\alpha=1/2$ it predicts a disk of radius $\sqrt2$ rather than the exact silver radius. This approximation therefore fails to reproduce the spectral support of the full random-matrix model.

The discrepancy can be seen without solving any resolvent equation. At $\alpha=0$, let
$A=C_A\otimes\mathbbm1_B/2$ and
$B=\mathbbm1_A\otimes C_B/2$, where the independent normalized GUE matrices have limiting variance one. These two centered tensor factors commute and satisfy
\begin{equation}
 \begin{split}
 \varphi(A^2)&=\varphi(B^2)=\frac14,\\
 \varphi(A^4)&=\varphi(B^4)=\frac18,\\
 \varphi(ABAB)&=\frac1{16}.
 \end{split}
 \label{eq:EC_app_mixed_tensor_moment}
\end{equation}
For freely independent centered variables the alternating moment would vanish. Here, however, the mixed terms survive, and
\begin{equation}
 \varphi[(A+B)^4]=\frac18+\frac18
 +6\left(\frac14\right)^2=\frac58,
 \label{eq:EC_app_fourth_moment}
\end{equation}
whereas replacing the two tensor legs by freely independent semicircular variables gives $1/2$. This leading-order difference is retained by the conditional matrix equation of Appendix~\ref{app:EC_MDE}.

\section{Numerical evaluation and convergence diagnostics}
\label{app:EC_numerics}

This appendix gives the numerical procedure used to evaluate the boundary integral in Sec.~\ref{sec:ensembleC_numerics}. The solid contours of Fig.~\ref{fig:ensembleC_generalalpha} are obtained without constructing finite random matrices. They are computed from Eq.~\eqref{eq:ensembleC_Fourier}, with $v_Y=1$.

\subsection{Physical branch and FFT evaluation}

For each candidate $z$ and contour parameter $q$, sample the ellipse $\lambda_q(\theta)$ from Eq.~\eqref{eq:EC_app_ellipse_param} at $N_\theta$ equispaced angles $\theta_j=2\pi j/N_\theta$. At each grid point $f=f_z(\lambda_q(\theta_j))$ solves the quadratic form of Eq.~\eqref{eq:EC_app_f},
\begin{equation}
 cf^2-(z-\lambda_q(\theta_j))f+1=0.
 \label{eq:EC_app_quadratic_f}
\end{equation}
For $c=0$, use $f=(z-\lambda_q)^{-1}$. Otherwise the physical quadratic root is selected by analytic continuation from the branch with $f\sim(z-\lambda)^{-1}$ at large spectral separation. It is tracked continuously around the ellipse. A discontinuity or failure to return to the initial root after one circuit signals a possible branch problem and invalidates that evaluation. Merely choosing a root independently at every node is insufficient near branch points.

The coefficients $F_{nk}$ are evaluated by an FFT on this angular grid. To incorporate the factor $(\sigma^2)^{n-1}$ in Eq.~\eqref{eq:ensembleC_Fourier}, it is convenient to use
\begin{equation}
 h_1(\theta)=f_z(\lambda_q(\theta)),\qquad
 h_{n+1}(\theta)=\sigma f_z(\lambda_q(\theta))h_n(\theta),
 \label{eq:EC_app_recursion}
\end{equation}
and denote the Fourier coefficients of $h_n$ by $\widehat h_{nk}$. The positive outer-order contribution then reads
\begin{equation}
\begin{split}
 T_n(z;q)&=\sum_{k\ge0}q^{2k}
 \left|\widehat h_{nk}-\frac{cq^2}{\sigma^2}
 \widehat h_{n,k+2}\right|^2,\\
 \Phi(z)&=\sum_{n\ge1}T_n(z;q).
\end{split}
 \label{eq:EC_app_Tn}
\end{equation}
The Fourier-index sum is truncated below the Nyquist index, leaving sufficient margin for the $k+2$ coefficient. Increasing $N_\theta$ checks both the Fourier cutoff and the angular quadrature. The recursion avoids separately computing very large powers of $f$ and $\sigma$.

\subsection{Truncation and two independent numerical checks}

All $T_n$ are non-negative. The partial sums of $\Phi$ therefore increase with $n$, and the convergence test must examine the positive tail rather than cancellation among successive orders. The implementation accumulates terms in blocks of 32 orders and requires several consecutive blocks to contribute less than a prescribed fraction of the accumulated sum. Representative strict settings use a relative block-tail tolerance $10^{-10}$ satisfied for four successive blocks and an upper cap of $1.5\times10^5$ outer orders. These numbers are numerical stopping parameters, not mathematical bounds on the infinite sum.

A small apparent tail is not sufficient to validate an integral. A second test changes the Cauchy contour, and a third test doubles the FFT grid. Denote by $\Phi(q,N_\theta)$ the result of a converged evaluation. The two comparison errors are
\begin{equation}
\begin{split}
 \epsilon_q&=\frac{|\Phi(q_2,N_\theta)-\Phi(q_1,N_\theta)|}{\mathcal N_q},\\
 \epsilon_N&=\frac{|\Phi(q_2,2N_\theta)-\Phi(q_2,N_\theta)|}{\mathcal N_N},
\end{split}
 \label{eq:EC_app_error_measures}
\end{equation}
where $\mathcal N_q$ and $\mathcal N_N$ are the respective
maxima of $1/v_Y$ and the two absolute values appearing in
the numerator of each comparison.
We use $q_2>q_1$ with both values below one. In the large-$\alpha$ calculations, a representative starting choice is $q_1=0.994$ and $q_2=0.997$, with $N_\theta=2048$ and an independent double-grid check. The strict calculations use tolerances of order $10^{-5}$ for both comparisons, allowing angular grids up to $16384$ points. If the FFT comparison is already small but the $q$ comparison fails, increasing $N_\theta$ alone is not expected to solve the problem. One may instead move both contours closer to $q=1$, provided the analytic branch remains admissible.

\subsection{Tracking the boundary and hitting unresolved points}

For a fixed polar direction $z=r e^{i\vartheta}$, or a rescaled radial parameterization of the same ray, the search begins with a point far enough outside the spectrum that a validated evaluation gives $v_Y\Phi(z)<1$. It then advances inward and seeks a bracket for the root of $v_Y\Phi(z)-1=0$. Accepted brackets are refined by bisection. The remaining quadrants follow from the reflection symmetries. The representative strict implementation uses 81 angular directions in the first quadrant, followed by the appropriate reflections.

A computed boundary point is retained only if its final integral evaluation passes the block-tail test, the $q$-deformation test, and the doubled-grid test, and if its boundary residual satisfies
\begin{equation}
 \epsilon_{\rm edge}=|v_Y\Phi(z)-1|\le10^{-3}.
 \label{eq:EC_app_boundary_residual}
\end{equation}
A trial at which the series has not converged is classified as 'unresolved'.

For $\alpha=5$ and $10$, angular directions near the top and bottom spectral tips become difficult to resolve. The analytic domain of $f_z$ restricts admissible integration contours, while the positive outer series can converge extremely slowly. 

\bibliography{literature}
\end{document}